\RequirePackage{ifpdf}
\ifpdf
\else
\PassOptionsToPackage{dvipdfmx}{graphicx}
\PassOptionsToPackage{dvipdfmx}{hyperref}
\fi

\documentclass[11pt,a4paper]{article}
\pdfoutput=1
\usepackage{jheppub}
\usepackage{amssymb,amsmath,amsthm,graphicx}
\usepackage{graphics, color}
\usepackage{subfigure}
\usepackage{latexsym}
\usepackage{bm}
\usepackage{epsfig}
\usepackage[utf8]{inputenc}

\def\ie{{\it i.e.} }
\def\eg{{\it e.g.} }

\begin{document}
\title{Gregory-Laflamme encounters Superradiance}
\author[1]{\'Oscar J.~C.~Dias,}
\author[2]{Takaaki~Ishii,}
\author[3]{Keiju~Murata,}
\author[4]{Jorge~E.~Santos,}
\author[5]{Benson~Way}

\affiliation[1]{STAG research centre and Mathematical Sciences, University of Southampton, University Road, Southampton SO17 1BJ, UK.}
\affiliation[2]{Department of Physics, Rikkyo University, Nishi-Ikebukuro, Tokyo 171-8501, Japan}
\affiliation[3]{Department of Physics, College of Humanities and Sciences, Nihon University, Sakurajosui, Tokyo 156-8550, Japan}
\affiliation[4]{DAMTP, Centre for Mathematical Sciences, University of Cambridge, Wilberforce Road, Cambridge CB3 0WA, UK}
\affiliation[5]{Departament de F\'{i}sica Qu\`{a}ntica i Astrof\'{i}sica, Institut de Ci\`{e}ncies del Cosmos\\
Universitat de Barcelona, Mart\'{i} i Franqu\`{e}s, 1, E-08028 Barcelona, Spain}
\emailAdd{ojcd1r13@soton.ac.uk}
\emailAdd{ishiitk@rikkyo.ac.jp}
\emailAdd{murata.keiju@nihon-u.ac.jp}
\emailAdd{jss55@cam.ac.uk}
\emailAdd{benson@icc.ub.edu}

\abstract{\noindent
We investigate the effect of superradiant scattering of gravitational perturbations on the stability of rotating black strings, focusing on the six dimensional equal-spinning Myers-Perry black string.  We find that rapidly rotating black strings are unstable to gravitational superradiant modes within a bounded range of string lengths.  The instability occurs because momentum along the string direction creates a potential barrier that allows for the confinement of superradiant modes.  Yet, five dimensional Myers-Perry black holes do not have stable particle orbits so, unlike other known superradiant systems, these black strings remain stable to perturbations with sufficiently high azimuthal mode number --- this is a `finite-$m$' superradiant instability.  For some parameters, this instability competes with the Gregory-Laflamme instability, but otherwise exists independently.  The onset of this instability is degenerate and branches to multiple steady-state solutions.  This paper is the first of a trilogy: in the next two, we construct two distinct families of rotating strings emerging from the superradiant onset (the `black resonator strings' and `helical black strings').  We argue that similar physics is present in 5-dimensional Kerr black strings, but not in $D>6$ equal-spinning Myers-Perry black strings.
}
\preprint{RUP-22-22}
\maketitle

\section{Introduction}

Despite its simplicity, the product spacetime of a Schwarzschild black hole and a circle, known as the Schwarzschild black string \cite{Horowitz:1991cd}, manages to capture a wealth of physical phenomena affecting black holes in higher dimensions.  Notable are the Gregory-Laflamme instability \cite{Gregory:1993vy,Gregory:1994bj}, its associated violation of the weak cosmic censorship conjecture \cite{Horowitz:2001cz,Lehner:2010pn,Figueras:2022zkg}, and its neighbouring solutions (non-uniform strings and localized black holes) which violate black hole uniqueness~(see e.g. \cite{Gubser:2001ac,Wiseman:2002zc,Sorkin:2004qq,Kudoh:2004hs,Sorkin:2006wp,Headrick:2009pv,Figueras:2012xj}).   It is now known that the same physics is ubiquitous among black holes with extended horizons, including those of particular interest to string theory and holography~\cite{Hanada:2007wn,Kawahara:2007fn}.

It is therefore natural to consider rotating black strings, as they are more generic generalizations of static black strings and can express behaviours that require the presence of angular momentum.  Of particular interest to us are superradiant instabilities.  Like the Gregory-Laflamme instability, the superradiant instability often also comes with violations of uniqueness with time-periodic black holes~\cite{Dias:2011at,Dias:2015rxy,Ishii:2018oms,Ishii:2019wfs,Ishii:2021xmn}, a possible violation of the weak cosmic censorship conjecture \cite{Dias:2011at,Dias:2015rxy,Niehoff:2015oga,Chesler:2018txn,Chesler:2021ehz}, and has wide applications to astrophysics, string theory, and holography~(see review \cite{Brito:2015oca}).

But the Gregory-Laflamme and superradiant instabilities are triggered by different mechanisms.  The Gregory-Laflamme instability is driven by horizon dynamics and the propensity for horizons to increase in area when there is a hierarchy of length scales.  By contrast, the superradiant instability occurs when waves are amplified by an ergoregion and then reflected back towards the horizon, leading to an exponentially growing instability. For black strings, momentum along the string direction produces an effective mass term, which can potentially provide enough confinement to trigger a superradiant instability that would not otherwise exist without this extra dimension \cite{Marolf:2004fya,Cardoso:2004zz,Cardoso:2005vk,Dias:2006zv}.  Indeed, such an instability exists for scalar field perturbations of Kerr black strings through this mechanism \cite{Cardoso:2004zz,Cardoso:2005vk}.

Yet, the story changes for higher dimensional rotating (Myers-Perry) black strings \cite{Myers:1986un,Hawking_1999}.  Indeed, unlike Kerr black strings, singly spinning\footnote{Recall that in five or higher dimensions there is more than one plane of rotation.} Myers-Perry black strings are stable to scalar field perturbations \cite{Cardoso:2005vk}. Furthermore, there are no stable bounded orbits around a wide variety of Myers-Perry black holes \cite{Tangherlini:1963,Frolov:2003en}\footnote{Some higher dimensional black holes admit stable bounded orbits, e.g. 6-dimensional singly spinning Myers-Perry black holes~\cite{Igata:2014xca} or 5-dimensional black rings~\cite{Igata:2010ye}.}, preventing particle-like waves from experiencing repeated amplification.  Consequently, no superradiant instability should exist for such black holes in the eikonal limit (i.e. point-particle limit), where the azimuthal wave number $m\to\infty$.  

Despite the lack of stable orbits, we demonstrate that rotating  black strings might, in certain cases, still be unstable to superradiance in the most universal sector: {\it gravitational} perturbations.  For simplicity, we will focus only on the six-dimensional equal-spinning Myers-Perry black string (\ie with equal angular momenta along two rotation planes) \cite{Gibbons:2004js,Gibbons_2005}, for which the associated gravitational perturbation equations reduce to ODEs. The five-dimensional equal-spinning Myers-Perry black hole lacks stable orbits~\cite{Kagramanova:2012hw}, which implies that its associated asymptotically $\mathcal{M}^{1,4}\times S^1$  black string cannot be unstable to superradiant perturbations in the eikonal limit $m\to\infty$.  Nevertheless, we will find that the instability exists for any finite $m$, with the unstable region approaching measure zero in parameter space as $m\to\infty$.  To our knowledge, this is the first system where such a {\it `finite-$m$' superradiant instability} is observed.\footnote{\label{foot:massiveScalar}Indeed, take the prototype example of a scalar field with mass $\mu$ in the Kerr black hole. In the limit of very large $\ell=m$ the growth rate of the superradiant instability scales as $e^{-4m\ln{(m)}}$ \cite{Eperon:2019viw}. Thus, for any value of $\mu$ or black hole rotation $a$, there is always a value of 
$m=m_\star\equiv \lceil \mu/\Omega_H \rceil $ above which the instability switches on. Thus all Kerr black holes are superradiant unstable to massive scalar field perturbations, no matter the value of $\Omega_H\leq \Omega_H^{\rm ext}$ and $\mu$ \cite{Eperon:2019viw}. On the other hand, unlike in the previous case, in our MP string we will find that as $m$ grows, the critical rotation (above which the superradiant instability can be present) increases and this value approaches extremality (i.e. the instability switches off) when $m\to \infty$.}  Because of this property, this instability is not expected to progress indefinitely to smaller and smaller length scales and lead to a violation of the weak cosmic censorship conjecture, unlike what was conjectured for `infinite-$m$' rotational superradiance seen in other systems \cite{Dias:2011at,Dias:2015rxy,Niehoff:2015oga}.

As mentioned earlier, the Gregory-Laflamme and superradiant instabilities are distinct physical phenomena and exist independently.  In this paper, we will also compare both of these instabilities.  Depending on the black hole and wave parameters, none, only one of them, or both instabilities can be present.  Where both instabilities coexist, we find that the Gregory-Laflamme instability typically has a much higher growth rate and we discuss possible scenarios for the time evolution of the system.

We further find linear evidence that this superradiant instability, much like those affecting rotating black holes, leads to a branch of new steady-state solutions   \cite{Basu:2010uz,Bhattacharyya:2010yg,Dias:2011tj,Dias:2011at,Dias:2013sdc,Cardoso:2013pza,Herdeiro:2014goa,Dias:2015rxy,Ishii:2018oms,Ishii:2020muv,Ishii:2021xmn} (these new solutions are often called `hairy' or `resonator' solutions).  Like the case for black holes \cite{Dias:2011tj,Herdeiro:2014goa,Dias:2015rxy,Ishii:2018oms}, the onset frequency (where $\mathrm{Im}(\omega)=0$) for superradiant modes has nonzero real part ($\mathrm{Re}(\omega)\neq0$), so the branching solutions are usually time-periodic, but not time independent nor axisymmetric.  
The onset of the instability in black strings can be {\it degenerate}, implying that multiple solutions branch from this onset (\ie more than a one-parameter family).  We will argue that it is natural to coin these solutions `{\it black resonator strings}' and `{\it helical black strings}', and we present perturbative evidence for their existence.  Then, in a sequel of two papers \cite{Dias:2022str,Dias:2023nbj}, we will construct them nonlinearly (using higher order perturbation theory and fully nonlinear numerical methods) and study their properties in detail.

The rest of the paper is structured as follows. Section \ref{sec:MP} reviews the equal-spinning Myers-Perry black string. Section~\ref{sec:GL} studies the onset and growth rate of the Gregory-Laflamme instability. We then investigate the superradiant instability in section~\ref{sec:Superradiance}, finding its onsets, associated instability regions and growth rates. We highlight that the onset of this instability provides perturbative evidence for the existence of novel (resonator and helical) black strings.   We end with a discussion of our results in Section~\ref{sec:Discussion}. In the appendices we argue that the gravitational superradiant instability is not present for spin 0 and 1 fields and that it also does not extend to $D>6$ equal-spinning black strings.

\section{Equal-spinning Myers-Perry black strings}\label{sec:MP}

The Myers-Perry black hole \cite{Myers:1986un,Hawking_1999} is an exact solution describing an asymptotically flat, rotating black hole in any dimension.  In odd dimensions and with all angular momenta set equal, the Myers-Perry black hole has an enhanced symmetry group permitting the solution to be written in a cohomogeneity-1 way (\ie depending nontrivially only on a single coordinate) \cite{Gibbons:2004js,Gibbons_2005}.  We will be working with the six-dimensional ($D=6$) black string given by the product of an equal-spinning five-dimensional Myers-Perry black hole and a circle. Hereafter, unless otherwise specified, the Myers-Perry (MP) black string refers to this specific solution.  Its metric can be written\footnote{The radial coordinate used here can be converted to the standard Boyer-Lindquist radial coordinate in \cite{Myers:1986un} through $r^2 \to r^2+a^2$.
}
\begin{eqnarray} \label{MPstring}
 && {\mathrm d}s^2 _{\rm MP \,string}= -\frac{F}{H}\,{\mathrm d}t^2 +\frac{{\mathrm d}r^2}{F} + r^2 \left[H\left( \frac{\sigma_3}{2} -\frac{\Omega}{H}\, {\mathrm d}t \right)^2+ \frac{1}{4}\left( \sigma_1^2 +\sigma_2^2\right)  \right] + \mathrm dz^2\,,
\end{eqnarray}
where
\begin{equation}\label{MPfns}
F(r)= 1-\frac{r_0^2}{r^2}+\frac{a^2r_0^2}{r^4}\,, \qquad H(r)=1+ \frac{a^2 r_0^2}{r^4}\,, \qquad
\Omega(r)=\frac{a \,r_0^2}{r^4}\,,
\end{equation}
and
\begin{equation}\label{Eulerforms}
\begin{split}
  \sigma_1 &= -\sin(2 \psi) \, \mathrm{d}\theta + \cos(2 \psi)\sin\theta \,\mathrm{d}\phi\ ,\\
  \sigma_2 &= \cos(2 \psi) \,\mathrm{d}\theta + \sin(2\psi)\sin\theta \,\mathrm{d}\phi\ ,\\
  \sigma_3 &= 2\,\mathrm{d}\psi + \cos\theta \,\mathrm{d}\phi
\end{split}
\end{equation}
are left-invariant one-forms that satisfy the Maurer-Cartan equation $\mathrm{d}\sigma_i = \frac{1}{2} \epsilon_{i j k} \sigma_j \wedge \sigma_k$. The Euler angles $(\theta,\phi,\psi)$ are coordinates on a squashed $S^3$ with ranges $0\leq \theta < \pi $, $0\leq \phi <2\pi$, and $0\leq \psi <2\pi$.  We take the coordinate $z\in(0,L)$ to be periodic with circle length $L$. Killing vectors of this spacetime are given by
\begin{equation}
 \begin{cases}
\partial_t, \: \partial_z, \:  \partial_\psi\,, & \\
 \xi_1 = \cos\phi\,\partial_\theta +
\frac{1}{2}\frac{\sin\phi}{\sin\theta}\,\partial_\psi -
\cot\theta\sin\phi\,\partial_\phi \,, & \\
\xi_2 = -\sin\phi\,\partial_\theta +
\frac{1}{2}\frac{\cos\phi}{\sin\theta}\,\partial_\psi -
\cot\theta\cos\phi\,\partial_\phi \,, & \\
\xi_3 = \partial_\phi \,, &
  \end{cases} \\
\label{lkv}
\end{equation}
and thus the isometry group of this spacetime is $\mathbb{R}_t \times U(1)_\psi \times SU(2)\times U(1)_z$, with
$\xi_i$ ($i=1,2,3$) being the generators of the $SU(2)$-symmetry group, $[\xi_i,\xi_j]=\epsilon_{ijk} \xi_k$.
The one-forms $\sigma_i$ are invariant under $\xi_j$, \emph{i.e.} $\pounds_{\xi_j}\,\sigma_i=0$ (where $\pounds$ stands for Lie derivative).

The solution has 3 dimensionful parameters: the mass radius $r_0$, the rotation parameter $a$, and the circle length $L$. The horizon radius $r_+$ is defined as the largest real root of $F$ which can be used to express the mass radius as $r_0=r_+/\sqrt{1-(a/r_+)^2}$. The energy, angular momentum, tension along $z$, temperature, angular velocity and entropy of the Myers-Perry string are, respectively:\footnote{
\label{footCanPsi}Note that we have defined $J$ and $\Omega_H$ with respect to a canonically normalised periodic variable $\psi\sim \psi+2\pi$, in accordance with the work of Myers and Perry \cite{Myers:1986un} (another option that is often taken would be $\psi\sim \psi+4\pi$); in these conventions, the superradiant factor (also known as resonant or synchronization factor) is given by $\omega-2m\Omega_H$. The non-standard factor of 2 reflects the fact that we have equal rotation along the two planes and we are using the period of $2\pi$ for $\psi$.}
\begin{equation}\label{ThermoMP}
\begin{split}
  E &= \frac{3\pi r_0^2 L}{8 G_6} \,, \qquad
  J =  \frac{\pi r_0^2 a L}{4 G_6} \,, \qquad T_z= \frac{\pi r_0^2 L}{8 G_6}  \,,  \\
  T_H &= \frac{1}{2 \pi  r_+} \frac{1-2 (a/r_+)^2}{\sqrt{1-(a/r_+)^2}}\,, \qquad S = \frac{L}{2G_6} \frac{\pi ^2 r_+^3}{\sqrt{1-(a/r_+) ^2}}\,, \qquad \Omega_H  = \frac{a}{r_+^2} \,, \\
\end{split}
\end{equation}
where $G_6$ is the six-dimensional Newton's constant.  In horizon radius units (equivalent to set $r_+\equiv1$), Myers-Perry black strings have two dimensionless parameters, which we will take for convenience to be $\Omega_H r_+$ and $k r_+$, where the wavenumber $k=2\pi/L$. Note that the temperature vanishes at $a/r_+=1/\sqrt{2}$, where the Myers-Perry black string is extremal with angular frequency $\Omega^{\mathrm{ext}}_H r_+=1/\sqrt{2}$.

We will show that the Myers-Perry black string is unstable to at least two sectors of gravitational perturbations. One is the familiar Gregory-Laflamme instability, first studied in the context of Schwarzschild black strings/branes  \cite{Gregory:1993vy,Gregory:1994bj} and then extended to rotating black strings in \cite{Kleihaus:2007dg,Dias:2009iu,Dias:2010eu,Dias:2010maa,Dias:2011jg}. The other is the superradiant instability whereby the Kaluza-Klein momentum along the string direction provides an effective mass term that confines superradiant bound states. This confinement mechanism for the instability was first proposed in \cite{Marolf:2004fya} and the perturbative analyses of the associated timescale, for Kerr strings, were done in \cite{Cardoso:2004zz,Dias:2006zv}. Numerical superradiant analysis of Kerr and single-spinning Myers-Perry black strings for scalar fields where also done in \cite{Cardoso:2005vk}.

Our linear mode stability analysis requires studying linearised gravitational perturbations.  Let $h_{AB}$ be a linear metric perturbation about the Myers-Perry black string.  We will choose to work in the traceless-transverse gauge defined by\footnote{\label{foot:indices}Greek indices $\mu,\nu,\ldots$ run only over the 5-dimensional coordinates $(t,r,\theta,\phi,\psi)$, capital Latin indices $A,B,\ldots$ run over these plus the extended direction $z$, and the lower case indices $i,j,k,\ldots$ run over the Euler angles $\{\theta,\phi,\psi\}$.}
\begin{equation}
\label{eqn:TTgauge}
h^A_{\phantom{A} A}=0\;, \qquad \nabla^A h_{AB} =0\,.
\end{equation}
In this gauge, the linearised Einstein equation reduces to
\begin{equation}
\label{Lichnerowicz}
(\Delta_L h)_{AB} \equiv -\nabla^C \nabla_C h_{AB} -2R_{A C B D} h^{C  D} = 0\,,
\end{equation}
where $\Delta_L$ is the Lichnerowicz operator, and all curvatures and covariant derivatives are computed with respect to the Myers-Perry black string background.

\section{The Gregory-Laflamme instability on Myers-Perry black strings}\label{sec:GL}

We now review the Gregory-Laflamme (GL) instability and study its properties in the (equal-spinning) Myers-Perry black string.  Generally, the Gregory-Laflamme instability is present when there are one or more directions where the horizon is elongated.  Intuitively, the Gregory-Laflamme instability is driven by the horizon's tendency to maximise its area (entropy) when there is a hierarchy of scales, while keeping the mass and angular momentum of the black hole fixed.  Often, the entropically preferred configuration with the same mass and angular momenta (and fixed Kaluza-Klein circle length) is a `localized' black hole with spherical horizon topology (or an array of such spherical black holes). Moreover, the dynamical transition from the black string into the localized black hole configuration necessarily comes with a change in horizon topology, which violates the weak cosmic censorship conjecture.  These are all fairly general qualitative features of gravitational systems with a hierarchy of horizon length scales, though specific details can differ.

In the case of the (equal-spinning) Myers-Perry black string, the two lengthscales are $r_+$ and $L$ and we expect the instability to occur at least for $L\gg r_+$. The resulting Gregory-Laflamme instability was already briefly studied in \cite{Kleihaus:2007dg,Dias:2010eu}, although in \cite{Dias:2010eu} more emphasis was given to the so-called ultraspinning instability of the equal angular momenta Myers-Perry black hole.\footnote{\label{footHarmonic}The ultraspinning instability affects rapidly rotating black holes in $D\geq6$ and their associated black strings in $D\geq7$, and do not affect the Myers-Perry black string studied here. We study the Gregory-Laflamme mode that is unstable for all values of the rotation, including in the Schwarszchild string limit. In the rotating Myers-Perry strings, there are axisymmetric Gregory-Laflamme instabilities that are only present for rotations above a critical value \cite{Dias:2010eu} that we do not discuss here.
}
We also note that the Gregory-Laflamme instabilities of single-spinning and arbitrarily spinning Myers-Perry black strings were also discussed in \cite{Dias:2009iu,Dias:2010maa} and \cite{Dias:2011jg}, respectively.

\subsection{Linear perturbations}\label{sec:GL1}

As stated above, we will work in the transverse-traceless gauge \eqref{eqn:TTgauge} which, in the present case, amounts to working in the gauge $h_{Az}=0$.  Then for convenience, we will only consider the components $h_{\mu\nu}$ with $(\mu,\nu)$ running only over the 5-dimensional coordinates $(t,r,\theta,\phi,\psi)$\footnote{To make contact with \cite{Dias:2010eu}, note that the transverse-traceless gauge conditions \eqref{eqn:TTgauge} reduce to $h^\mu{}_{\mu}=0\;, \nabla^\mu h_{\mu\nu} =0$ and the linearised equation reads $ (\Delta_L h)_{\mu\nu} \equiv -\nabla^\rho \nabla_\rho h_{\mu \nu} -2R_{\mu \rho \nu \sigma} h^{\rho  \sigma} = -k^2 h_{\mu\nu}$. 
}.

The Gregory-Laflamme perturbations of the Myers-Perry string \eqref{MPstring} that we investigate have the form $\mathrm ds^2=\mathrm ds^2_\mathrm{\rm MP \,string}+ \delta \mathrm ds^2_\mathrm{\hbox{\tiny GL}} $ with
\begin{align}\label{GLpert}
\delta \mathrm ds^2_\mathrm{\hbox{\tiny GL}}&=h_{\mu\nu}  \mathrm dX^\mu \mathrm dX^\nu \\
&= e^{{\rm i} kz} e^{-{\rm i}\,\omega \, t} \Bigg(
-\frac{F}{H}\,\Big(Q_1 {\mathrm d}t +2 Q_6{\mathrm d}r \Big){\mathrm d}t
+ \frac{{\mathrm d}r}{F} \Big(Q_2 {\mathrm d}r+Q_7 \sigma_3 \Big)  \nonumber\\
& \qquad+ r^2 \left[ H\left( \frac{\sigma_3}{2} -\frac{\Omega}{H} {\mathrm d}t \right)
\bigg( Q_3\left( \frac{\sigma_3}{2} -\frac{\Omega}{H} {\mathrm d}t \right)-2\frac{\Omega}{H} Q_5 {\mathrm d}t  \bigg)
+ \frac{Q_4}{4}\left(\sigma_1^2+\sigma_2^2\right)  \right]  \Bigg)\,, \nonumber
\end{align}
where $Q_i$ ($i=1,\ldots,7$) are unknown functions of $r$ and we have introduced the mode frequency $\omega$ and the Kaluza-Klein momentum or wavenumber $k=2\pi/L$.

Note that this ansatz does not encompass the most general set of gravitational perturbations that trigger a Gregory-Laflamme instability.  We have set the azimuthal wave number $m$ associated to the Killing vector $\partial_\psi$ of the background to zero, as the Gregory-Laflamme instability is strongest in the axisymmetric sector of perturbations.  We have also not included any $\theta$ and $\phi$ dependence.\footnote{These can be introduced by decomposing the perturbations into charged harmonics on $\mathbb{C}\mathrm{P}^{1} \simeq S^2$ \cite{Hoxha:2000jf,Kunduri:2006qa,Martin:2008pf,Dias:2010eu}. There are scalar, vector and tensor harmonics on $\mathbb{C}\mathrm{P}^{1}$ but only the scalar describes Gregory-Laflamme instabilities \cite{Hoxha:2000jf,Kunduri:2006qa,Martin:2008pf,Dias:2010eu,Ishii:2018oms}.
}
Gregory-Laflamme instabilities with such angular dependence (i.e. with $m\neq 0$) \cite{Dias:2010eu} only exist above a critical rotation and typically have a smaller growth rate.
The modes that we study exist even without rotation.

The linearised Einstein equation \eqref{Lichnerowicz} and the gauge conditions \eqref{eqn:TTgauge} can be reduced to a system of three second order equations for $\{Q_2,Q_3,Q_5\}$. This can be accomplished as follows. Using the gauge conditions \eqref{eqn:TTgauge}, $Q_1$ and $Q_4$ can be expressed algebraically in terms of the other functions and their derivatives.  The linearised Einstein equation \eqref{Lichnerowicz} then reduces to a second order equation for $Q_2$ and four first order equations for $Q_{3,5,6,7}$. We can further use the first order equations for $Q_3$ and $Q_5$ to get two algebraic equations for $Q_6$ and $Q_7$, which finally converts the equations to a system of three second order equations for $\{Q_2,Q_3,Q_5\}$ (when $\omega=0$, we get a similar system of 3 coupled ODEs but with $Q_6=Q_7=0$).

We now discuss boundary conditions.  The asymptotic behaviour of the functions is $Q_i|_{r \to\infty}\sim e^{\pm \sqrt{k^2-\omega^2} \,r}$. We keep only the exponentially decaying solution to preserve Kaluza-Klein asymptotics ($\mathcal{M}^{1,4}\times S^1$). In more detail, the Gregory-Laflamme modes decay as $Q_{2,3,5}|_{r\to\infty}\sim e^{-\sqrt{k^2-\omega^2}\,r}r^{\,\alpha_{2,3,5}}$ where $\alpha_i$ are constants.\footnote{For time dependent perturbations ($\omega\neq 0$) one has $\alpha_2= \alpha_3= -3/2$ and $\alpha_5= 3/2$ while for the zero mode problem ($\omega= 0$) one has $\alpha_2=-5/2$, $ \alpha_3=-3/2$ and $\alpha_5= 3/2$.} At the horizon, we have $Q_i |_{\cal H}\sim (r-r_+)^{\pm {\rm i} \,\frac{\omega}{4 \pi  T_H}}$. We keep the solution with the minus sign as it is the one that is regular in ingoing Eddington-Finkelstein coordinates. In more detail, the Gregory-Laflamme modes behave as $Q_{2,3,5}|_{r\to r_+}\sim (r-r_+)^{\beta_{2,3,5}- {\rm i} \,\frac{\omega}{4 \pi  T_H}}$ with $\beta_{2}=-1$ and $\beta_{3}=\beta_{5}=0$.\footnote{\label{footEF}The Myers-Perry string background is smooth across the event horizon if we use ingoing Eddington-Finkelstein coordinates $(v,r,\theta,\phi,\Psi)$ given by:  $v=t+\int H^{1/2}F^{-1} \,\mathrm{d}r\,,
\Psi=\psi+2\int \Omega\, H^{-1/2}F^{-1} \,\mathrm{d}r\,$. For time-dependent perturbations ($\omega\neq 0$) one has $\beta_{2}=-1$ and $\beta_{3}=\beta_{5}=0$, while for the zero mode problem ($\omega= 0$) one has $\beta_{2}=\beta_{3}=0$ and $\beta_{5}=1$.}

For numerics, it is convenient to introduce a radial coordinate
\begin{equation}\label{CompactCoord}
y=\sqrt{1-\frac{r_+}{r}},
\end{equation}
which ranges $y\in (0,1)$ with the horizon $r=r_+$ being at $y=0$ and the asymptotic region $r\to\infty$ at $y=1$.

We will work in units of the horizon radius (equivalent to $r_+\equiv 1$), where the dimensionless quantities are given by
\begin{equation}\label{dimAux}
\widetilde{a}=a/r_+\,, \quad \widetilde{L}=L/r_+ \quad \widetilde{\Omega}_H=\Omega_H r_+\,, \quad  \widetilde{T}_H=T_H r_+\,, \quad \widetilde{k}=k r_+\,, \quad  \widetilde{\omega}=\omega r_+\,.
\end{equation}
In these units, the Myers-Perry black string can be parametrised by $\widetilde\Omega_H$ and $\widetilde k$.

Our numerical study of the Gregory-Laflamme instability proceeds in two parts, each of which require separate numerical computations.  First, while scanning through $\widetilde\Omega_H$, we obtain a critical wavenumber $\widetilde k=\widetilde k_{(0)}|_{\hbox{\tiny GL}}$ that yields the onset zero mode value $\widetilde\omega=0$.  Second, we solve the general problem for $\widetilde\omega\neq0$ while scanning through parameter space $(\widetilde\Omega_H, \widetilde k$).

In the zero mode case with $\omega=0$, the boundary conditions can be enforced by the field redefinitions
\begin{align}\label{PertVariables:GLzeroMode}
&Q_2(y)=e^{-\frac{\widetilde{k}}{1-y^2}} \left(1-y^2\right)^{5/2} q^{(0)}_1(y)\,, \nonumber\\
& Q_3(y)=e^{-\frac{\widetilde{k}}{1-y^2}} \left(1-y^2\right)^{3/2} q^{(0)}_2(y)\,, \\
& Q_5(y)=e^{-\frac{\widetilde{k}}{1-y^2}} \left(1-y^2\right)^{-3/2} y^2 q^{(0)}_3(y)\,, \nonumber
\end{align}
and requiring that the $q^{(0)}_i$'s are smooth functions. From the equations of motion, it follows that the $q^{(0)}_i$'s now obey Neumann conditions at the horizon and Robin conditions at infinity.

With $\widetilde\omega=0$, the critical wavenumber $\widetilde k=\widetilde k_{(0)}|_{\hbox{\tiny GL}}$ appears as a quadratic eigenvalue problem.  For any given rotation below extremality 
$0 \leq \widetilde\Omega_H \leq 1/\sqrt{2}$, there is a unique solution with real, positive $\widetilde k=\widetilde k_{(0)}|_{\hbox{\tiny GL}}$, which we identify as the critical wavenumber for the onset of the Gregory-Laflamme instability. This gives us the onset curve $\widetilde k_{(0)}|_{\hbox{\tiny GL}}(\widetilde\Omega_H)$.  Note there is also a trivial $\widetilde k=0$ solution.

As a result of this analysis, we will find that unstable Myers-Perry black strings lie in the rectangular region of parameter space given by $0 \leq \widetilde\Omega_H \leq 1/\sqrt{2}$ and $0<\widetilde k<\widetilde k_{(0)}|_{\hbox{\tiny GL}}(\widetilde\Omega_H)$.

With the zero mode solution in hand, we can then proceed to the second part of our numerical study, this time with $\widetilde\omega\neq0$.  For this case, boundary conditions can be imposed by the field redefinitions
\begin{align}\label{PertVariables:GL}
&Q_2(y)=e^{-\frac{\sqrt{\widetilde{k}^2-\widetilde{\omega}^2}}{1-y^2}} \left(1-y^2\right)^{3/2} y^{-2-\frac{{\rm i}\,\widetilde{\omega}}{2\pi \widetilde{T}_H}}q_1(y)\,, \nonumber\\
& Q_3(y)=e^{-\frac{\sqrt{\widetilde{k}^2-\widetilde{\omega}^2}}{1-y^2}} \left(1-y^2\right)^{3/2} y^{-\frac{{\rm i}\,\widetilde{\omega}}{2\pi \widetilde{T}_H}}q_2(y)\,, \\
& Q_5(y)=e^{-\frac{\sqrt{\widetilde{k}^2-\widetilde{\omega}^2}}{1-y^2}} \left(1-y^2\right)^{-3/2} y^{-\frac{{\rm i}\,\widetilde{\omega}}{2\pi \widetilde{T}_H}}q_3(y)\,, \nonumber
\end{align}
again with the requirement that the $q_i$'s are smooth functions.

Now, the aim is to compute the frequency $\widetilde\omega$ for any given $\widetilde k$ and $\widetilde\Omega_H$ in the unstable region.  This is a non-polynomial eigenvalue problem in $\widetilde\omega$, which can be approached by using a Newton-Raphson algorithm with the zero mode solution as an initial seed. To fix the normalization of the eigenvalue problem, we set $Q_5|_{y=0}=1$. (See e.g. \cite{Cardoso:2013pza,Dias:2015nua} for details on using the Newton-Raphson method for non-polynomial eigenvalue problems).

Since Gregory-Laflamme modes have purely imaginary frequencies, ${\rm Re}\,\widetilde\omega=0$, it is convenient to redefine the frequency as $\widetilde\Gamma={\rm i}\,\widetilde\omega$, which reduces the domain of the eigenvalue problem from the complex plane to the real line. Further note that one then has $e^{-\frac{\sqrt{\widetilde{k}^2-\widetilde{\omega}^2}}{1-y^2}}=e^{-\frac{\sqrt{\widetilde{k}^2+\widetilde{\Gamma}^2}}{1-y^2}}$ and asymptotic decaying modes should exist for any $\widetilde k$, although only long wavelengths with $\widetilde{k}\leq \widetilde{k}_{(0)}|_{\hbox{\tiny GL}}$, should be unstable.

Numerically, we discretise the differential equations using pseudo-spectral methods on a Chebyshev grid.  As expected for pseudospectral method, our results converge exponentially with the number of grid points. All the frequencies that will be presented in our plots are accurate up to at least the eighth decimal digit. Moreover, our results match those of \cite{Dias:2010eu}\footnote{The numerical codes used were developed independently.}.

\subsection{Gregory-Laflamme unstable region and growth rates}\label{sec:GL2}

\begin{figure}[t]
\centering
\includegraphics[width=.4\textwidth]{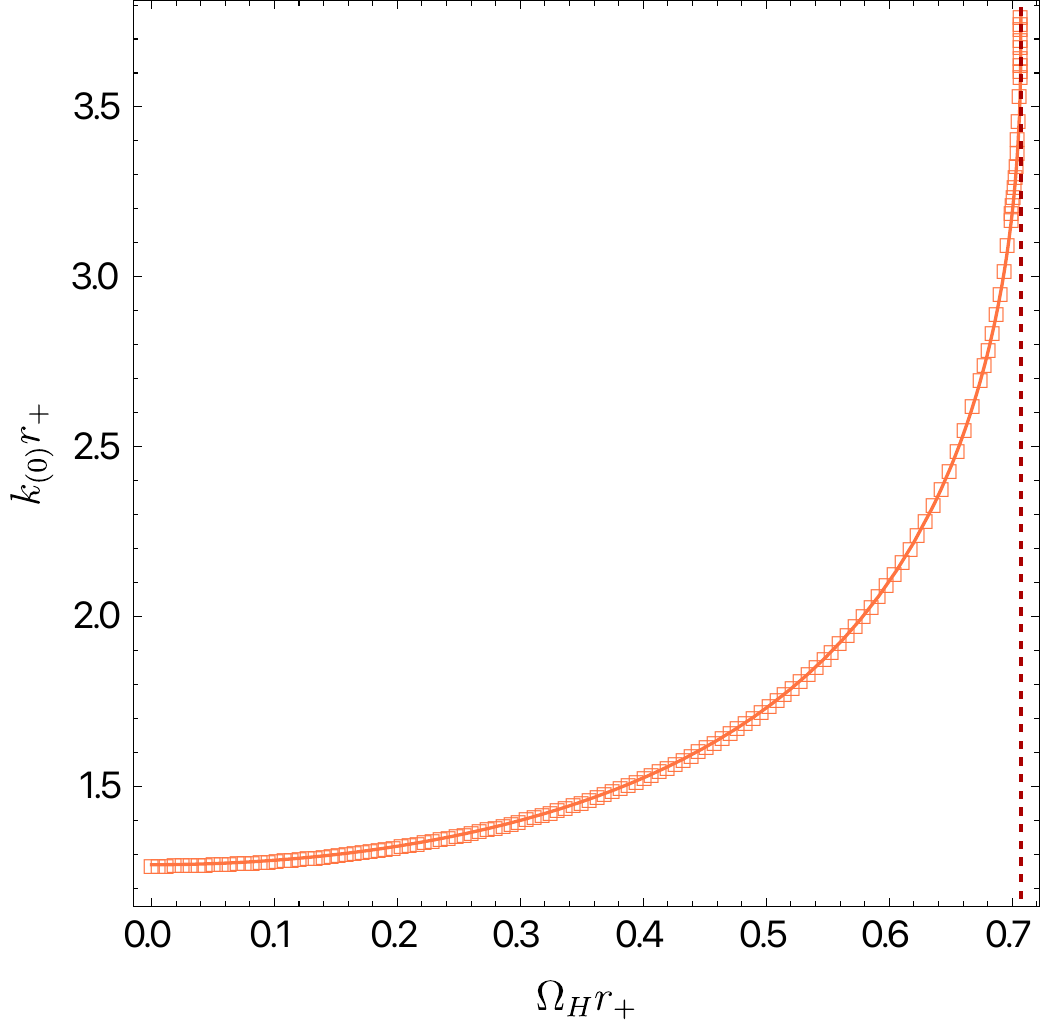}
\hspace{1cm}
\includegraphics[width=.51\textwidth]{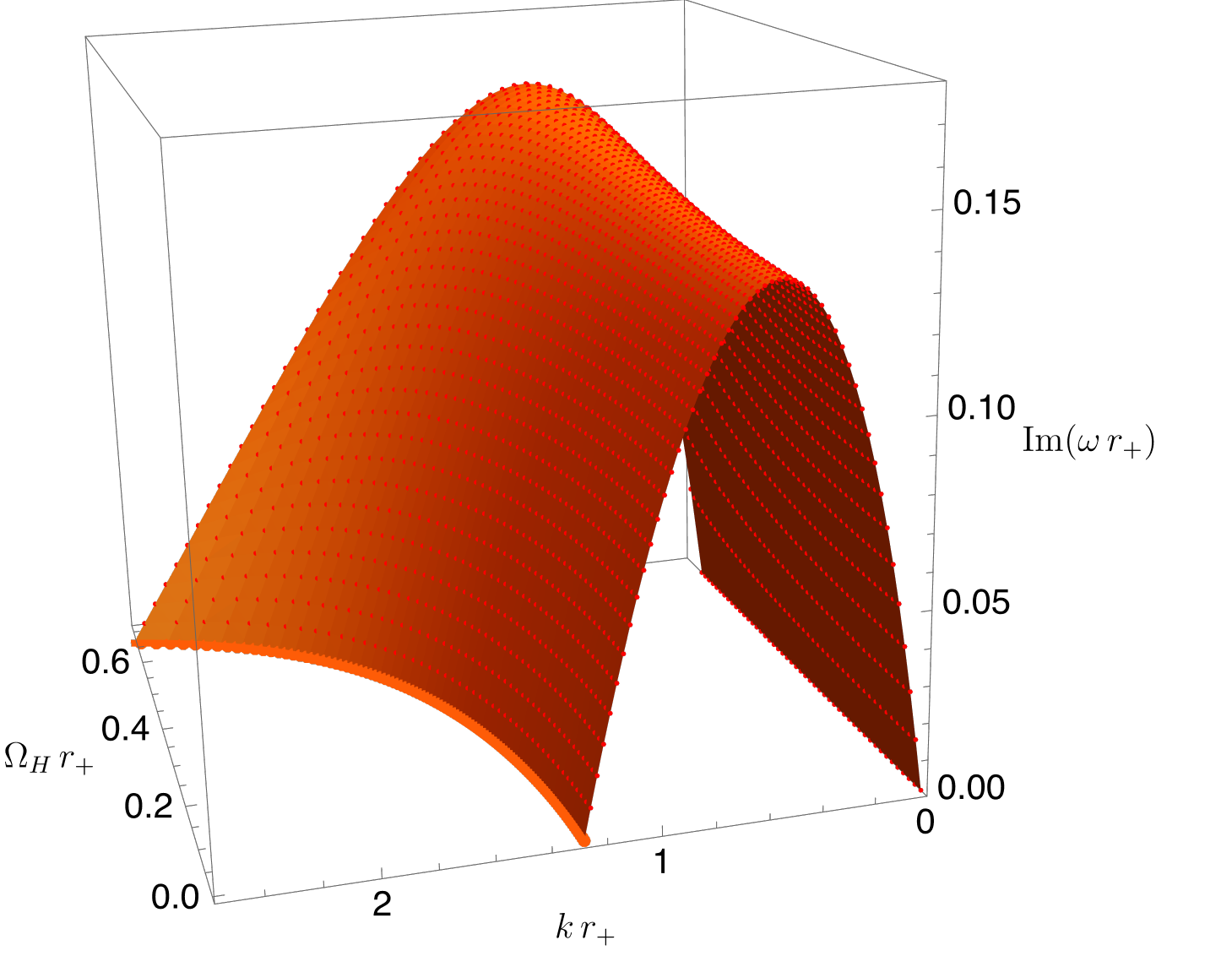}
\caption{Onset mode wavenumber (left) and growth rate (right) of the axisymmetric ($m=0$) Gregory-Laflamme instability for Myers-Perry black strings.
In the left panel, unstable modes have $\widetilde{k} <\widetilde{k}_{(0)}$ (\ie $\widetilde L>2\pi/\widetilde{k}_{(0)}$ are unstable). The vertical dashed line marks extremality with $\widetilde{\Omega}_H=1/\sqrt{2}\simeq 0.707107$.
Critical wavenumbers are $\widetilde{k}_{(0)}  \simeq 1.268916$
at $\widetilde\Omega_H =0$ \cite{Gregory:1993vy} and $ \widetilde{k}_{(0)}\simeq 3.767342$ at
$\Omega_H \simeq 0.999996\,\Omega^{\mathrm {ext}}_H \simeq 0.707104$ (extrapolates to $ \widetilde{k}_{(0)} \simeq 3.783201$ at $\Omega_H=\Omega^{\mathrm {ext}}_H$).
In the right panel, we only show solutions up to $\widetilde{\Omega}_H\sim 0.6 <1/\sqrt{2}$.  In this range, the growth rate increases with increasing $\widetilde{\Omega}_H$ and fixed $\widetilde k$. For higher $\widetilde{\Omega}_H$, the growth rate falls: see discussion of right panel of Fig.~\ref{Fig:timescaleGL2} for further details). All modes here have purely imaginary frequency, ${\rm Re}\, \omega=0$.
} \label{Fig:GLmodes}
\end{figure}

We now present our results for the Gregory-Laflamme instability. The zero mode wavenumber $\widetilde{k}_{(0)}\equiv \widetilde{k}_{(0)}|_{\hbox{\tiny GL}}$ and growth rates $\mathrm{Im} \, \widetilde{\omega}$ are shown in the left and right panels of Fig.~\ref{Fig:GLmodes}, respectively. 

From the left panel, one sees that the critical wavenumber $\widetilde k_{(0)}$ exists for all values of the angular velocity up to extremality, $0 \leq \widetilde\Omega_H\leq 1/\sqrt{2}$.   For zero rotation, the unstable region is $0< \widetilde k \lesssim 1.268916$, which agrees with the instability of the Schwarzschild-Tangherlini black string \cite{Gregory:1993vy}. As rotation is increased towards extremality, the unstable region grows to $0< \widetilde k \lesssim 3.783$. In this sense, we can state that rotating black strings are more prone to the Gregory-Laflamme instability.

The right panel of Fig.~\ref{Fig:GLmodes} displays the Gregory-Laflamme growth rate $\widetilde{\Gamma}={\rm Im}(\omega r_+)$ across the dimensionless parameters $\widetilde\Omega_H$ and $\widetilde k$. The onset zero mode curve with $\widetilde{k}=\widetilde{k}_{(0)}|_{\hbox{\tiny GL}}(\widetilde{\Omega}_H)$ and ${\rm Im}\,\widetilde{\omega}=0$, is also displayed in the right panel of Fig.~\ref{Fig:GLmodes} as the orange curve. The fact that the growth rate computation returns ${\rm Im}\,\widetilde{\omega}=0$ along the same line $\widetilde{k}=\widetilde{k}_{(0)}|_{\hbox{\tiny GL}}(\widetilde{\Omega}_H)$ as the independent zero mode numerical code is a non-trivial check of our computations.

\begin{figure}[th]
\centering
\includegraphics[width=.45\textwidth]{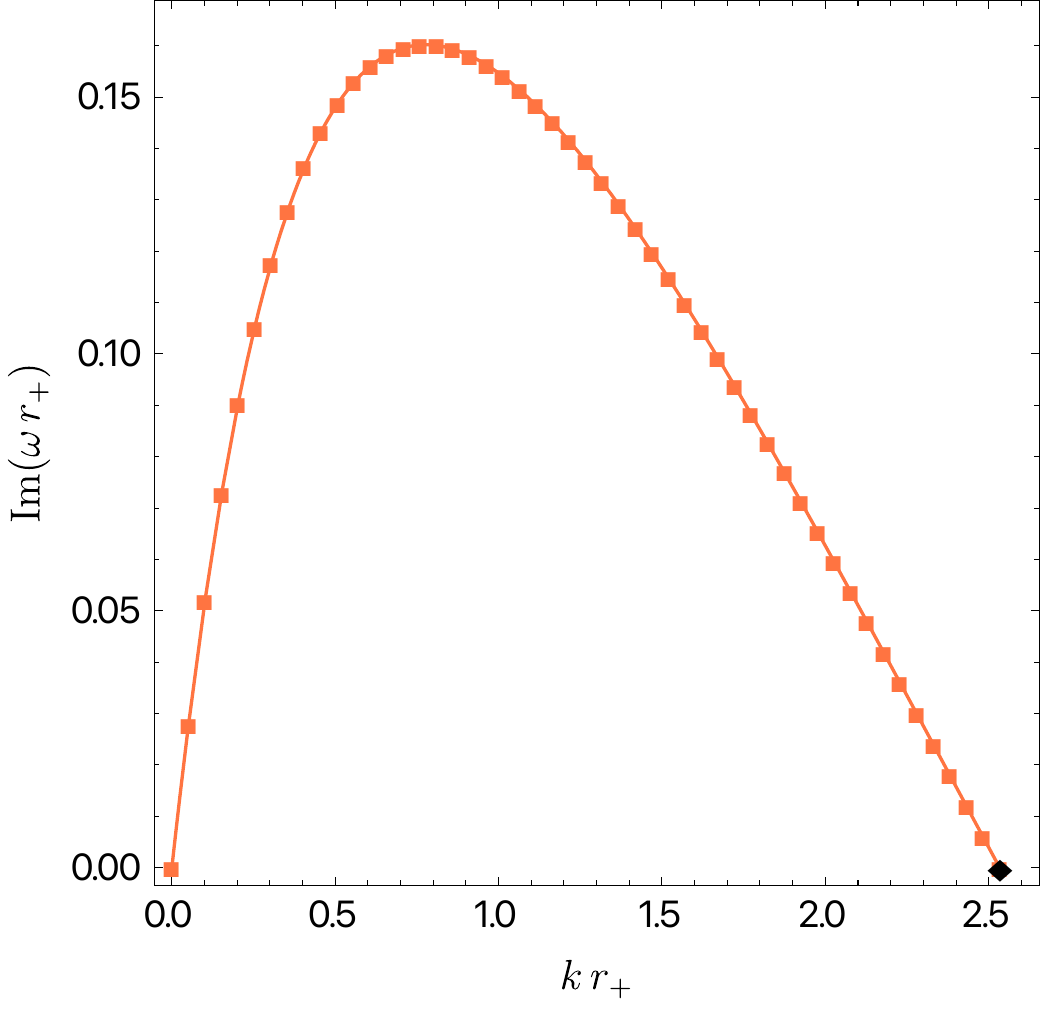}
\hspace{1cm}
\includegraphics[width=.45\textwidth]{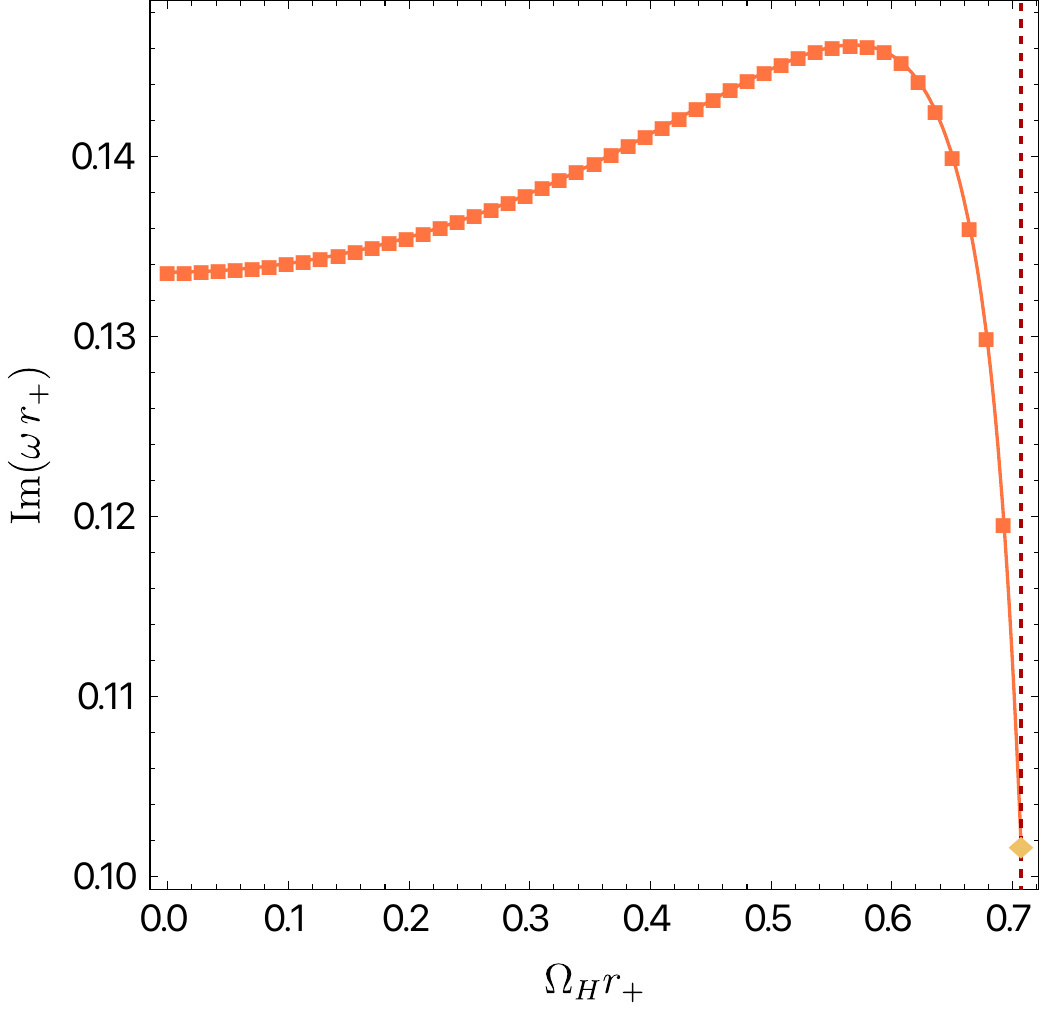}
\caption{Growth rate for the Gregory-Laflamme instability of the Myers-Perry black string with fixed $\widetilde{\Omega}_H=0.66$ (left), and with $\widetilde{k}=\frac{1}{2}\,\widetilde{k}^{\mathrm{GL}}_{(0)}$ (right). Note that in the right plot, $\widetilde{k}^{\mathrm{GL}}_{(0)}$ changes with $\widetilde{\Omega}_H$ according to Fig.~\ref{Fig:GLmodes}, so $\widetilde{k}$ varies as well.  The yellow diamond in the right panel at $(\widetilde{\Omega}_H,\widetilde{\omega})\simeq (1/\sqrt{2},0.101756)$ is the extrapolated value of the growth rate at extremality (vertical dashed line).
} \label{Fig:timescaleGL2}
\end{figure}

From Fig.~\ref{Fig:GLmodes}, we see that Myers-Perry black strings are Gregory-Laflamme unstable for the whole rotation range $0 \leq  \widetilde{\Omega}_H\leq 1/\sqrt{2}$ whenever the wavenumber is below critical $0<\widetilde k<\widetilde k_{(0)}|_{\hbox{\tiny GL}}(\widetilde\Omega_H)$, that is, whenever the length is higher than the zero mode critical length, $\widetilde{L}\geq \widetilde{L}_{(0)}|_{\hbox{\tiny GL}}$. For most range of parameters (aside from those close to extremality which we will discuss later in Fig.~\ref{Fig:timescaleGL2}), the growth rate of the Gregory-Laflamme instability increases as the rotation increases at fixed $\widetilde L$. Moreover, for a given $\widetilde{\Omega}_H$ the maximum growth rate is higher than the one for the Schwarzschild-Tangherlini black string. Thus, in this sense, we see that the Gregory-Laflamme instability can get stronger with rotation.

For completeness, in the left panel of Fig.~\ref{Fig:timescaleGL2} we focus our attention on Myers-Perry black strings with $\widetilde\Omega_H=0.66$ and describe how the Gregory-Laflamme growth rate changes as  $0\leq \widetilde{k} \leq \widetilde{k}_{(0)}|_{\hbox{\tiny GL}} \simeq 2.533565$. This is a curve that is qualitatively similar to that of the Schwarzschild string. On the right panel, we select Myers-Perry strings that have a length that is twice as large as the critical onset length, $\widetilde{L}=2 \widetilde{L}_{(0)}|_{\hbox{\tiny GL}}$, for each rotation (note $\widetilde{L}$ is not constant as $\widetilde{\Omega}_H$ changes). We see that starting at zero rotation, the growth rate first increases, attains a maximum, and then decreases substantially in a short window of angular velocity as extremality is approached. Note that in the right panel of Fig.~\ref{Fig:GLmodes} this sharp decrease is not visible because we do not show the solutions too close to extremality where the fall occurs, i.e. we just display solutions up to $\widetilde{\Omega}_H\sim 0.6 <1/\sqrt{2}$.

Summarizing our results, as long as we are not too close to extremality, adding rotation tends to increase the parameter range of unstable black strings, as well as increase the unstable growth rate. In both these senses, the Gregory-Laflamme instability tends to get stronger with added rotation.  Our results are qualitatively consistent with previous results found in \cite{Dias:2010eu}.

\section{The superradiant instability of Myers-Perry black strings}\label{sec:Superradiance}

In this section, we will show that the Myers-Perry black string is unstable to gravitational superradiance and compare the associated  unstable region and growth rate to those of the Gregory-Laflamme instability.

Superradiant instabilities require two primary ingredients.  The first is the existence of superradiant scattering, \ie an amplification mechanism of incident waves that enter an ergoregion. This is present in the Myers-Perry string.  The second is a confining mechanism whereby the amplified waves are reflected back to the ergoregion and can continue to extract energy and angular momenta until they back-react on the spacetime geometry.  At the outset, it is unclear whether the Myers-Perry black string satisfies this second criterion.

Heuristically, non-trivial Kaluza-Klein modes (i.e. modes with frequency $\omega$ and momentum $k$ along the string direction) decay asymptotically as $e^{-\sqrt{k^2-\omega^2}\,r}$, with the momentum $k$ providing an effective mass proportional to $\sqrt{k^2-\omega^2}$.  This effective mass creates a potential barrier along the radial direction that might provide enough of a confining mechanism, as first proposed in \cite{Marolf:2004fya} (see also \cite{Cardoso:2004zz,Cardoso:2005vk,Dias:2006zv}). Though this argument applies for generic (integer spin) perturbations, the fact is that no such instability has been found for scalar and Maxwell fields on the (equal-spinning) Myers-Perry black string.  Indeed, we have explicitly checked that there is no superradiant instability for these perturbations and, actually, we can provide an analytical argument for the absence of the instability as discussed in Appendix~\ref{app:ScalarMaxwell}.  For the scalar field on the single spinning Myers-Perry black string, \cite{Cardoso:2005vk} also observed the lack of a superradiant instability.  Furthermore, the lack of stable orbits around 5-dimensional equal-spinning Myers-Perry black holes also strongly suggests that there should not be a superradiant instability in the associated six-dimensional Myers-Perry black string in the eikonal limit (where the azimuthal mode number $m\to\infty$).

Yet, despite the above negative results, we will nevertheless find a superradiant instability for {\it gravitational} perturbations in six-dimensional equal-spinning Myers-Perry black strings, for any {\it finite} azimuthal number $m$. Perhaps surprisingly, this result appears to be valid only for $D=6$ equal-spinning Myers-Perry black strings. Indeed, in Appendix~\ref{app:higherD}, we provide an argument suggesting the absence of the instability in $D=2N+4$ dimensional equal-spinning Myers-Perry black strings for $N\geq 2$ (i.e. for $D\geq 8$). So $D=6$ is very special.

\subsection{Wigner D-matrices}\label{sec:Wigner}

Unlike the Gregory-Laflamme instability, superradiant instabilities that are sourced by rotation require non-axisymmetric ($m\neq 0$) modes.  It is known on fairly general grounds that superradiant modes satisfy $\mathrm{Re}(\omega)<2 m\Omega_H$, where $m$ is the azimuthal mode number, in this case for the coordinate $\psi$ (recall footnote~\ref{footCanPsi}).  We seek special sectors of perturbations that are decoupled and induce the superradiant instability in the Myers-Perry black string.

To describe decoupled perturbations in these conditions, we use the Wigner D-matrices $D^{j}_{\ell'\ell}(\theta,\phi,\psi)$. We ask the reader to see \cite{Ishii:2021xmn} for a detailed review of the Wigner D-matrices in the context of black hole perturbations and here we just highlight key aspects that are fundamental for our study.

Let us define the angular momentum operators $L_i\equiv \mathrm{i} \xi_i$ and $R_3\equiv \mathrm{i} \frac{1}{2}\partial_\psi$. They satisfy the commutation relations, $[L_i,L_j]=\mathrm{i} \epsilon_{ijk}L_k$ and $[L_i,R_3]=0$. We can take the set of mutually commutative operators $(L^2,L_3,R_3)$ where $L^2\equiv L_1^2+L_2^2+L_3^2$. Wigner D-matrices are defined as eigenfunctions of these operators:
\begin{equation}
 L^2 D^{j}_{\ell'\ell} = j(j+1)D^{j}_{\ell'\ell}\ ,\quad
 L_3 D^{j}_{\ell'\ell} = \ell' D^{j}_{\ell'\ell}\ ,\quad
 R_3 D^{j}_{\ell'\ell} = \ell D^{j}_{\ell'\ell}\ .
\end{equation}
The quantum numbers can take the values
\begin{equation}
 j=0,\frac{1}{2},1,\frac{3}{2},\dots\ ,\qquad \ell',\ell=-j,-j+1,\dots, j\ .
 \label{eq:jota}
\end{equation}
There are convenient formulae for the derivatives of Wigner D-matrices:
\begin{equation}\label{WignerFomulae}
\begin{split}
\partial_\theta D_\ell &= -\frac{\mathrm{i}}{2}(\epsilon_{-} e^{-2\mathrm{i}\psi} D_{\ell-1}+\epsilon_{+} e^{2\mathrm{i}\psi} D_{\ell+1})\,,\\
\partial_\phi D_\ell &= -\mathrm{i} \ell \cos\theta D_\ell +\frac{1}{2}\epsilon_- \sin\theta \, e^{-2\mathrm{i}\psi} D_{\ell-1} -\frac{1}{2}\epsilon_{+} \sin\theta \, e^{2\mathrm{i}\psi} D_{\ell+1} \,,\\
\partial_\psi D_\ell &= -2\,\mathrm{i} \ell D_\ell\,,
\end{split}
\end{equation}
where $\epsilon_{\pm} \equiv \sqrt{(j \mp \ell)(j \pm \ell +1)}$ and the common indices $(j,\ell')$ were suppressed to simplify the notation. We ask the reader to see \cite{Hu:1974hh,Ishii:2020muv} for a derivation of these results.

For the following perturbation analysis, it is convenient to introduce 1-forms $\sigma_\pm$ defined by
\begin{equation}
\sigma_{\pm}=\frac{1}{2}(\sigma_1\mp \mathrm{i} \, \sigma_2)=\frac{1}{2}e^{\mp 2\,\mathrm{i}\,\psi}(\mp \mathrm{i} \,\mathrm{d}\theta+\sin\theta \mathrm{d}\phi)\,,
\label{sigmapm}
\end{equation}
where $\sigma_{1,2}$ were defined in \eqref{Eulerforms}.  The one-forms $\sigma_\pm$ are ``eigen 1-forms'' of $R_3$, \ie $\pounds_{R_3} \sigma_\pm = \pm \sigma_\pm$.

The eigenvalue of $-R_3$ is the azimuthal mode number $m$. For example, we say that $D^j_{\ell'\ell}$ has azimuthal mode number $m=-\ell$. We can also define the azimuthal mode number for tensors. For example, tensors $D^j_{\ell'\ell}\sigma_-^2$, $D^j_{\ell'\ell}\sigma_+\sigma_-$ and $D^j_{\ell'\ell}\sigma_+^2$ have $m=-\ell+2, -\ell, -\ell-2$, respectively.

\subsection{Linear perturbations}\label{sec:Superradiance1}

This time we focus our attention on a traceless-transverse perturbation of \eqref{MPstring} that has the form $\mathrm ds^2= \mathrm ds^2_\mathrm{\rm MP \,string}+ \delta \mathrm ds^2_{\hbox{\tiny SR}} $ where
\begin{align}\label{SRpert}
\delta \mathrm ds^2_{\hbox{\tiny SR}}&= h_{MN}\mathrm dx^{M}\mathrm dx^{N}=r^2 e^{{\rm i} \, k \, z} e^{-{\rm i}\,\omega \, t} D^j_{\ell' \ell} \,Q(r) \sigma_{-}^2\quad\text{with}\quad \ell'=\ell=-j\,, \nonumber \\
& = \frac{1}{4}e^{{\rm i} \, k \, z} e^{-{\rm i}\,\omega \, t} e^{{\rm i}\,(m-2)\,(\phi+2\psi)} \cos^{2(m-2)}\left(\frac{\theta}{2}\right) \,Q(r)\left(\sigma_1^2-\sigma_2^2+2\,{\rm i}\,\sigma_1\,\sigma_2\right)
\end{align}
which describes a superradiant perturbation with a charged scalar $\mathbb{C}\mathrm{P}^{1}$ harmonic dependence in $\psi$ corresponding to azimuthal quantum number
$m=j+2$, and thus $m\in\{2,5/2,3,7/2,\ldots\}$; see~\eqref{eq:jota}. Since this is the maximum azimuthal wavenumber for a given $j$, this perturbation decouples from other perturbations and it is the simplest disturbance that permits superradiance in the Myers-Perry black string.

The perturbation \eqref{SRpert} is trivially in the traceless-transverse gauge \eqref{eqn:TTgauge} and reduces the Lichnerowicz equation \eqref{Lichnerowicz} to a second order ODE for $Q(r)$:
\begin{equation}\label{EOMsuper}
\left(r^3 F Q' \right)'+r\left[ \frac{H r^2}{F}\left(\omega -2 m \frac{\Omega}{H}\right)^2-\frac{4m^2}{H}+4m-k^2 r^2 \right]Q=0\,.
\end{equation}
Using the tortoise coordinate, $dr_\ast = \sqrt{H}/F dr$, and redefining the perturbation variable as $\Phi=(rH)^{1/4}Q$, \eqref{EOMsuper} can be rewritten in the Schr\"{o}dinger form
\begin{equation}
 \frac{d^2 \Phi}{dr_\ast^2} - V \Phi = 0\ ,
\label{Scheq}
\end{equation}
where its potential reads:
\begin{equation}
V\equiv \frac{F}{H} k^2 -\left(\omega -2 m \frac{\Omega}{H}\right)^2 + \frac{4 F}{r^2 H} \left[ (m-1)m - m^2 \left(1 - \frac{1}{H}\right)\right]
+\frac{1}{r^{3/2} H^{1/4}}  \frac{\mathrm{d}^2}{\mathrm{d} r_\ast^2}\left(r^{3/2} H^{1/4}\right)\ .
\end{equation}
Near infinity, the Schr\"{o}dinger potential $V$ behaves as
\begin{equation}
 V|_{r\to \infty}\simeq  k^2-\omega^2 +\left[4m(m-1)+\frac{3}{4}-\frac{k^2r_+^2}{1-\Omega_H^2r_+^2}\right]\frac{1}{r^2} + \mathcal{O}\left(r^{-4}\right).
\label{Vinf}
\end{equation}
Note that the coefficient of $1/r^2$ becomes negative for a sufficiently large $k$.
This suggests that the gravitational perturbation might get confined near the horizon (inside the ergoregion) and thus there might be room for a superradiant instability~\cite{Cardoso:2005vk} (note that in the scalar field perturbation case, there is no Schr\"odinger potential barrier that confines the superradiant modes which is consistent with the fact that there is no superradiant instability; see appendix~\ref{app:ScalarMaxwell}). Moreover, for $m\to \infty$, the $1/r^2$ term in \eqref{Vinf} becomes positive. This is related to the fact that there are no stable particle orbits in the five-dimensional Myers-Perry black hole  \cite{Tangherlini:1963,Frolov:2003en}. So if the superradiant instability happens to be present for the Myers-Perry black string at finite $m$, it should definitely shut down in the eikonal ($m\to\infty$) limit.

The analysis above suggests the possible existence of a superradiant instability for finite $m$, which we now attempt to find. To search for possible unstable modes of \eqref{Scheq}, we need to impose the relevant physical boundary conditions. At infinity, an asymptotic analysis of the ODE \eqref{EOMsuper} indicates the existence of two independent solutions: $Q\sim r^{-3/2} e^{\pm \sqrt{k^2-\omega^2}\,r}$. We discard the exponentially growing mode to preserve the Kaluza-Klein $\mathcal{M}^{1,4}\times S_z^1$ asymptotics. Near the horizon, the ODE has again two independent solutions: $Q\sim (r-r_+)^{\pm {\rm i} \,\frac{\omega -2m\Omega_H}{4 \pi  T_H}}$. We impose boundary conditions that retain only the solution with the minus sign to maintain regularity in ingoing Eddington-Finkelstein coordinates (see footnote~\ref{footEF}).  Introducing, as we have for the Gregory-Laflamme instability, the compact radial coordinate $y=\sqrt{1-\frac{r_+}{r}}$ and working in units of $r_+$ as in \eqref{dimAux}, these boundary conditions can be imposed by setting
\begin{equation}\label{PertVariable}
Q(y)=e^{-\frac{\sqrt{\widetilde{k}^2-\widetilde{\omega}^2}}{1-y^2}}\left(1-y^2\right)^{3/2}  \Big[\big(2-y^2\big)y^2\Big]^{-{\rm i}\,\frac{\widetilde{\omega}-2m \widetilde{\Omega}_H}{4 \pi \widetilde{T}_H}}q(y)
\end{equation}
and requiring that $q$ is a smooth function. Smoothness of $q$ and the equations of motion now yield a Neumann condition at the horizon and a Robin condition at infinity for $q(y)$. Given parameters $\widetilde{\Omega}_H$ and $\widetilde{k}$, our ODE and its boundary conditions describe a quadratic eigenvalue problem for $\widetilde{\omega}$.

To find the onset of the instability we will set $\widetilde\omega=2m\widetilde\Omega_H$ and $\widetilde k=\widetilde k_{(0)}$ (recall footnote~\ref{footCanPsi}). The above discussion on boundary conditions still holds in this particular case, but the problem becomes a quadratic eigenvalue problem for $\widetilde{\kappa}\equiv  \sqrt{\widetilde{k}_{(0)}^2-4m^2 \widetilde{\Omega}_H^2}$.

Our numerical computation now proceeds as follows.  As before, we find the onset curve first, and then find the growth rates.  We first fix $m\in \{2,5/2,3,7/2,\ldots\}$ (we will start with $m=2$ and later consider higher $m$). Then, scanning over $\widetilde{\Omega}_H$, we attempt to find the onset modes by setting $\widetilde\omega=2m \widetilde\Omega_H$ (the condition for the onset of superradiance), relabelling $\widetilde k=\widetilde k_{(0)}$, then solving the corresponding quadratic eigenvalue problem for $\widetilde{\kappa}$ on a Myers-Perry black string family, which also ultimately gives us a value for $\widetilde k_{(0)}$. We expect real, positive solutions for $\widetilde{\kappa}$ to exist for a range of dimensionless rotations $\widetilde{\Omega}_H|_c<\widetilde{\Omega}_H<\widetilde{\Omega}^{\mathrm{ext}}_H=1/\sqrt2$, which gives us an onset curve $\widetilde{k}_{(0)}|_{\hbox{\tiny SR}}(\widetilde\Omega_H)$ for this range of $\widetilde{\Omega}_H$.

Having found the critical $\widetilde{k}_{(0)}(\widetilde{\Omega}_H)$ for a given $m$, we then expect that for values of $\widetilde{k}$ just critical value $\widetilde{k}\lesssim\widetilde{k}_{{0}}$, the mode will be unstable with frequency ${\rm Im} \,\widetilde{\omega}>0$.  We can find this frequency by solving the aforementioned quadratic eigenvalue problem for $\widetilde{\omega}$.  We can then continue to vary $\widetilde\Omega_H$  and $\widetilde k$ to explore the regions where ${\rm Im}\,\widetilde{\omega}>0$.  Numerically, we again rely on pseudospectral methods with a Chebyshev grid and use a Newton-Raphson method to track modes as we vary parameters. We follow the strategy described in \cite{Cardoso:2013pza,Dias:2015nua} to solve eigenvalue problems using  Newton-Raphson, fixing the normalization by setting $q|_{y=0}=1$.

\subsection{Superradiant unstable regions and growth rates}\label{sec:Superradiance2}

\begin{figure}[b]
\centering
\includegraphics[width=.477\textwidth]{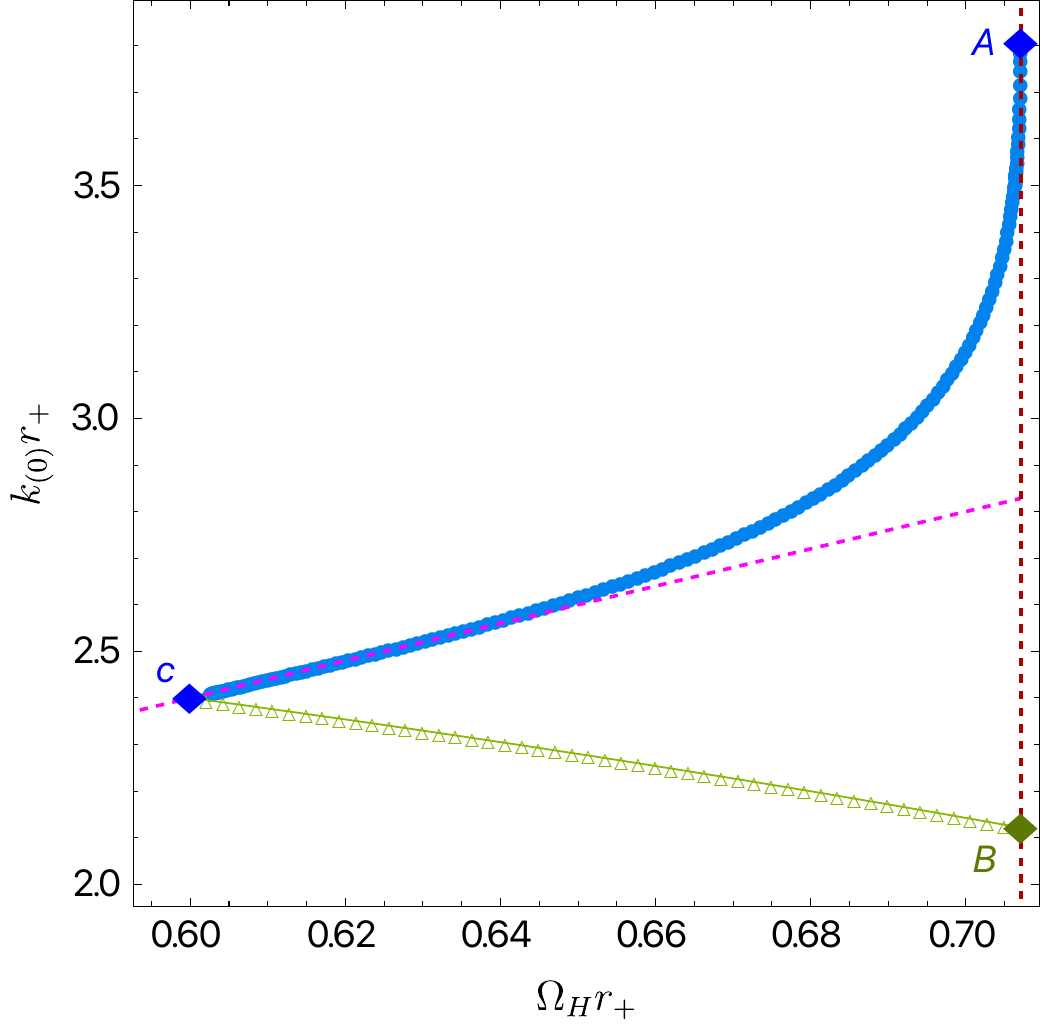}\hspace{1cm}
\includegraphics[width=.441\textwidth]{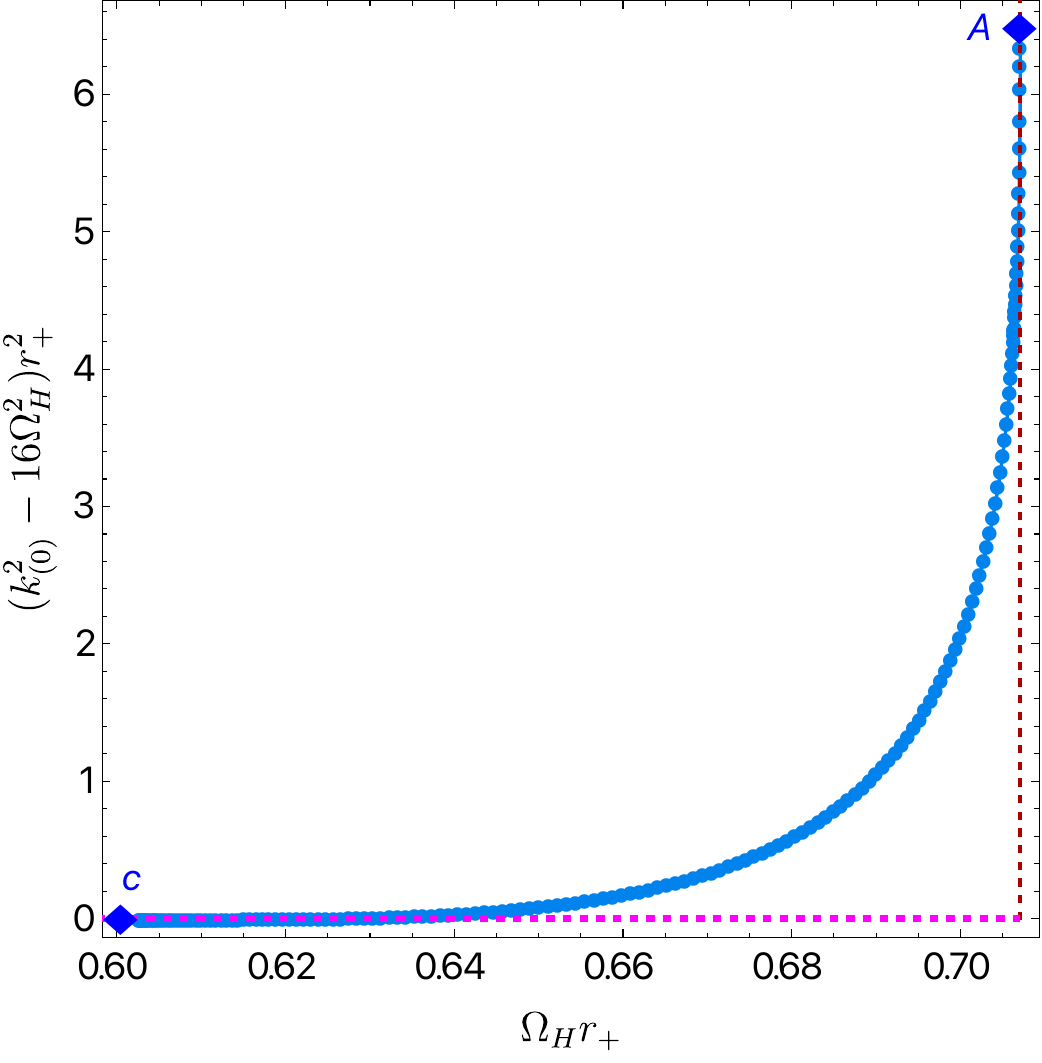}
\caption{The blue disk curves with endpoints $A$ and $c$ corresponds to the $m=2$ onset mode (with $\widetilde\omega=2m\widetilde\Omega_H$) of the superradiant instability of the Myers-Perry black string. This curve exists in the window $  \widetilde{\Omega}_H|_c \leq \widetilde{\Omega}_H \leq  \widetilde{\Omega}_H|_A$ where $ \widetilde{\Omega}_H|_c=3/5$ and $ \widetilde{\Omega}_H|_A\equiv\widetilde{\Omega}^{\mathrm{ext}}_H= 1/\sqrt{2}$ is the angular velocity at extremality (the vertical dashed line identifies this extremality).
{\bf Left:} 
Myers-Perry black strings are superradiant unstable in the triangular region $ABc$ bounded by the shown curves. Besides the onset curve ($Ac$), we also show the green curve $\widetilde{k}_{\star}(\widetilde{\Omega}_H)$, which describes modes with ${\rm Re\, \widetilde{\omega}}=\widetilde{k}$ (and ${\rm Im}\,\widetilde{\omega}=0$). This green curve $\widetilde{k}_{\star}(\widetilde{\Omega}_H)$ begins at point $c$ where it intersects the magenta dashed line $\widetilde{k}=2 m \widetilde{\Omega}_H$ (with $m=2$), and ends at point $B$ where $\widetilde{\Omega}_H=\widetilde{\Omega}^{\mathrm{ext}}_H= 1/\sqrt{2}$. For reference, point $A$ has $(\widetilde{\Omega}_H, \widetilde{k}_{(0)}, \widetilde{L}_{(0)})_{A}\simeq (1/\sqrt{2},3.805857,1.650925)$, point $B$ has $(\widetilde{\Omega}_H, \widetilde{k}_{(0)}, \widetilde{L}_{(0)})_{B}= (1/\sqrt{2},3/\sqrt{2},2\sqrt{2}\pi/3)$, and point $c$ has $( \widetilde{\Omega}_H, \widetilde{k}_{(0)}, \widetilde{L}_{(0)})_{c}= (3/5, 12/5, 5\pi/6)$, where $\widetilde{L}_{(0)}=2\pi/\widetilde{k}_{(0)}$.
{\bf Right:} This plot demonstrates that the onset curve $Ac$ satisfies $\widetilde{k}_{(0)}^2-4m^2 \widetilde{\Omega}_H^2\geq 0$, which implies that these onset modes decay exponentially in the radial direction.
}
 \label{Fig:zeroMode}
\end{figure}

Let us discuss the results in detail for $m=2$ before proceeding with higher values of $m$.  We begin with the critical superradiant onset curve $\widetilde{k}_{(0)}(\widetilde{\Omega}_H)$, where $\widetilde\omega=2m\widetilde\Omega_H$ (and is critical because ${\rm Im}\,\widetilde{\omega}=0$).  This curve is the blue $cA$ curve in Fig.~\ref{Fig:zeroMode}, and exists for a range of rotation $\widetilde{\Omega}_H|_c<\widetilde{\Omega}_H<\widetilde{\Omega}^{\mathrm{ext}}_H=1/\sqrt2$.  Within this range, wavenumbers just below critical $\widetilde{k}\lesssim\widetilde{k}_{{0}}$ are unstable.

For reasons that will later become clear, we wish to demonstrate that the onset modes at $\widetilde{k}_{(0)}(\widetilde{\Omega}_H)$ decay exponentially in the radial direction (as argued previously when discussing the asymptotic boundary condition of the problem).  Since, by the known asymptotic behaviour and our boundary conditions, the solutions have a falloff given by $e^{-\sqrt{k^2-\omega^2}\,r}$ and this onset curve $cA$ satisfies $\widetilde\omega=2m\widetilde\Omega_H$, we simply need to show that $\widetilde k^2-4m^2\widetilde\Omega_H^2>0$. We can easily see from the right panel of Fig.~\ref{Fig:zeroMode} that this condition is satisfied within the given range $\widetilde{\Omega}_H|_c<\widetilde{\Omega}_H\leq\widetilde{\Omega}^{\mathrm{ext}}_H=1/\sqrt2$, and appears to be marginal, $\widetilde{k}\to 2m\widetilde{\Omega}_H$, at $\widetilde{\Omega}_H=\widetilde{\Omega}_H|_c$.

\begin{figure}[th]
\centering
\includegraphics[width=.505\textwidth]{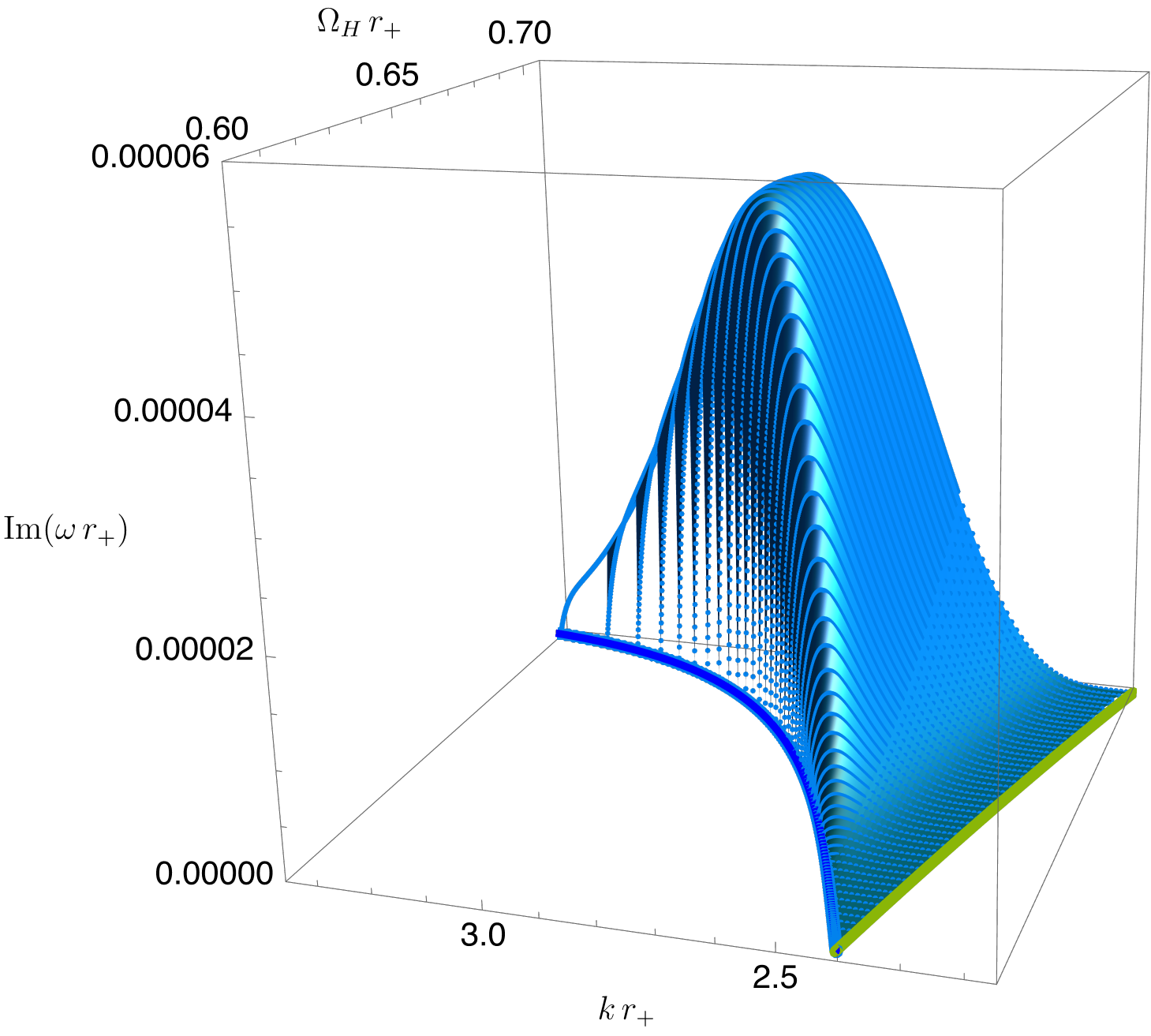}
\hspace{-0.2cm}
\includegraphics[width=.49\textwidth]{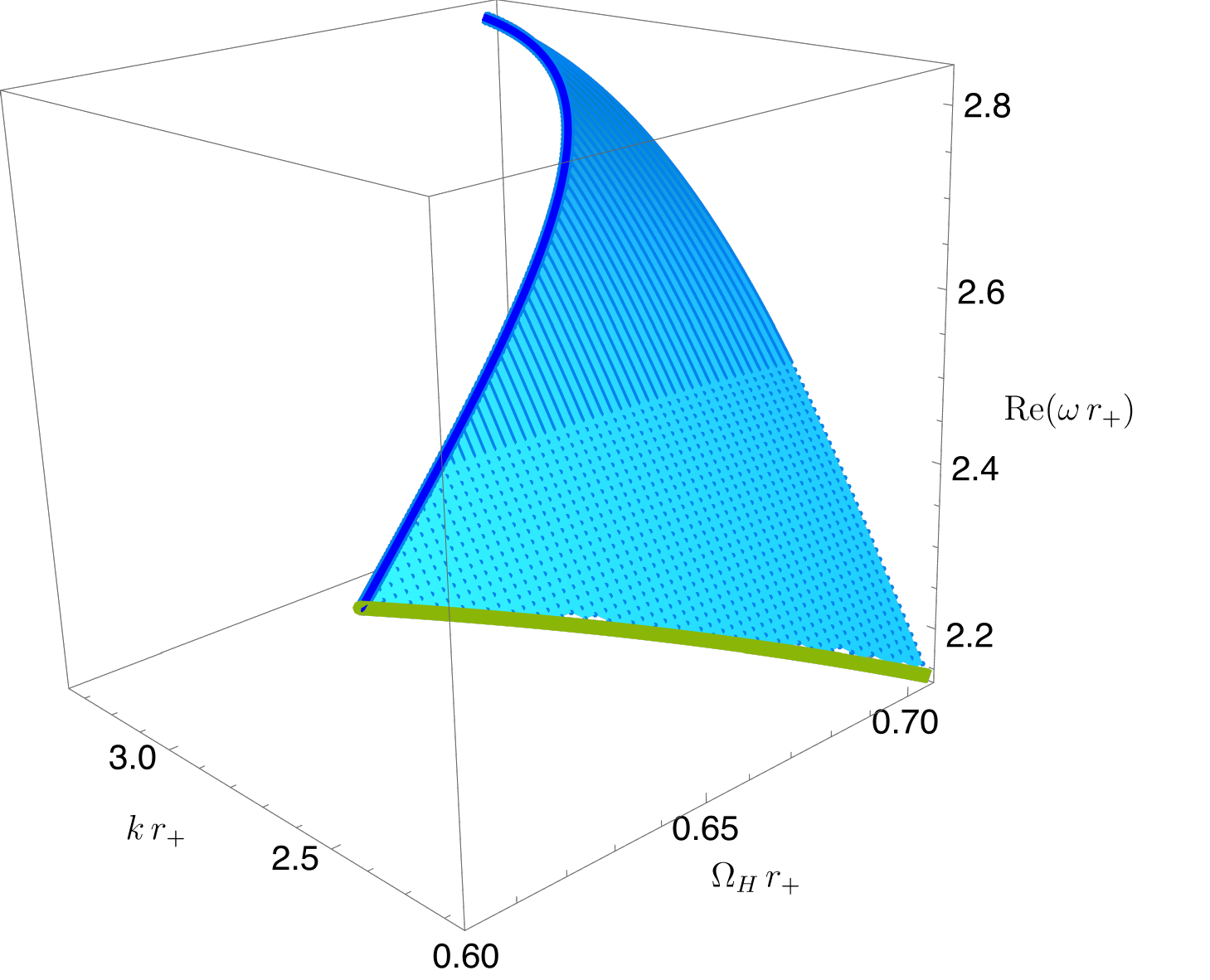}
\caption{Frequency for $m=2$ unstable superradiant modes, as a function of parameters of Myers-Perry black strings. The instability is present (\ie ${\rm Im}\,\widetilde{\omega}>0$) for $  \widetilde{\Omega}_H|_c \leq  \widetilde{\Omega}_H \leq  1/\sqrt{2}$, with $ \widetilde{\Omega}_H|_c =3/5$, and $  \widetilde{k}_{\star}(\widetilde{\Omega}_H)\leq \widetilde{k}\leq \widetilde{k}_{(0)}(\widetilde{\Omega}_H)$. The critical curve $\widetilde{k}_{(0)}(\widetilde{\Omega}_H)$ (and ${\rm Im}\,\widetilde{\omega}=0$) was also computed independently in Fig.~\ref{Fig:zeroMode} and we  show it as a continuous blue line in the ${\rm Im}\,\widetilde{\omega}=0$ plane. On the other hand, the  green curve represents the line  $\widetilde{k}_{\star}(\widetilde{\Omega}_H)$ as defined parametrically in  \eqref{CutoffHighm}. It meets the merger blue curve at $ \widetilde{\Omega}_H|_c =3/5$. The ${\rm Im}\,\widetilde{\omega}=0$ plane matches the left panel of Fig.~\ref{Fig:zeroMode}.
}
\label{Fig:timescaleSuper}
\end{figure}

After finding this onset curve $\widetilde{k}_{(0)}(\widetilde{\Omega}_H)$, we can then explore the range of parameters where an instability exists (\ie where ${\rm Im}\,\widetilde{\omega}>0$).  The real and imaginary parts of $\widetilde\omega$ for parameters with such an instability are shown in Fig.~\ref{Fig:timescaleSuper}.  We see that the unstable region in $(\widetilde\Omega_H,\widetilde k)$ parameter space is triangular, and lies between the blue onset curve $\widetilde{k}_{(0)}(\widetilde{\Omega}_H)$, extremality $\widetilde\Omega_H=\widetilde\Omega^{\mathrm{ext}}_H=1/\sqrt2$, and a new green curve which we will call $\widetilde{k}_{\star}(\widetilde{\Omega}_H)$.  This unstable triangular region is the same as the triangular region $ABc$ shown in the left panel of Fig.~\ref{Fig:zeroMode}.
The blue and green curves (both with ${\rm Im}\,\widetilde{\omega}=0$) of Fig.~\ref{Fig:timescaleSuper} correspond, respectively, to the blue onset curve $cA$ and green $cB$ curve of Fig.~\ref{Fig:zeroMode}. On other hand the extremal $AB$ curve in~ Fig.~\ref{Fig:zeroMode} corresponds to a curve with ${\rm Im}\,\widetilde{\omega}>0$ at $\widetilde{\Omega}_H=\widetilde{\Omega}_H^{\rm ext}$.

Let us now explain the physical origin of this new green curve $\widetilde{k}_{\star}(\widetilde{\Omega}_H)$.  As we have mentioned earlier, superradiant instabilities require two ingredients: amplification from an ergoregion, and a confining mechanism.  The blue onset curve $\widetilde{k}_{(0)}(\widetilde{\Omega}_H)$ we have found satisfies the superradiant condition $\widetilde\omega=2m \widetilde\Omega_H$, which corresponds to the place where the amplification mechanism shuts off.  It is natural, then, to suggest that the green curve $\widetilde{k}_{\star}(\widetilde{\Omega}_H)$ corresponds to where the confining mechanism shuts off.  This occurs when the exponential fall-off of $e^{-\sqrt{k^2-\omega^2}\,r}$ becomes marginal with $\widetilde k=\widetilde\omega$ (recall that the curve $\widetilde{k}_{\star}(\widetilde{\Omega}_H)$ already has $\mathrm{Im}\,\widetilde\omega=0$).

We have already seen evidence for this idea from the right panel of Fig.~\ref{Fig:zeroMode} and the discussion a few paragraphs ago, where we show that the modes on the onset $Ac$ curve $\widetilde{k}_{(0)}(\widetilde{\Omega}_H)$ have this exponential fall-off and that this fall-off becomes marginal at $\widetilde{\Omega}_H=\widetilde{\Omega}_H|_c$, which is precisely the location where the onset curve $\widetilde{k}_{(0)}(\widetilde{\Omega}_H)$ intersects this new green $cB$ curve $\widetilde{k}_{\star}(\widetilde{\Omega}_H)$.

\begin{figure}[th]
\centering
\includegraphics[width=.485\textwidth]{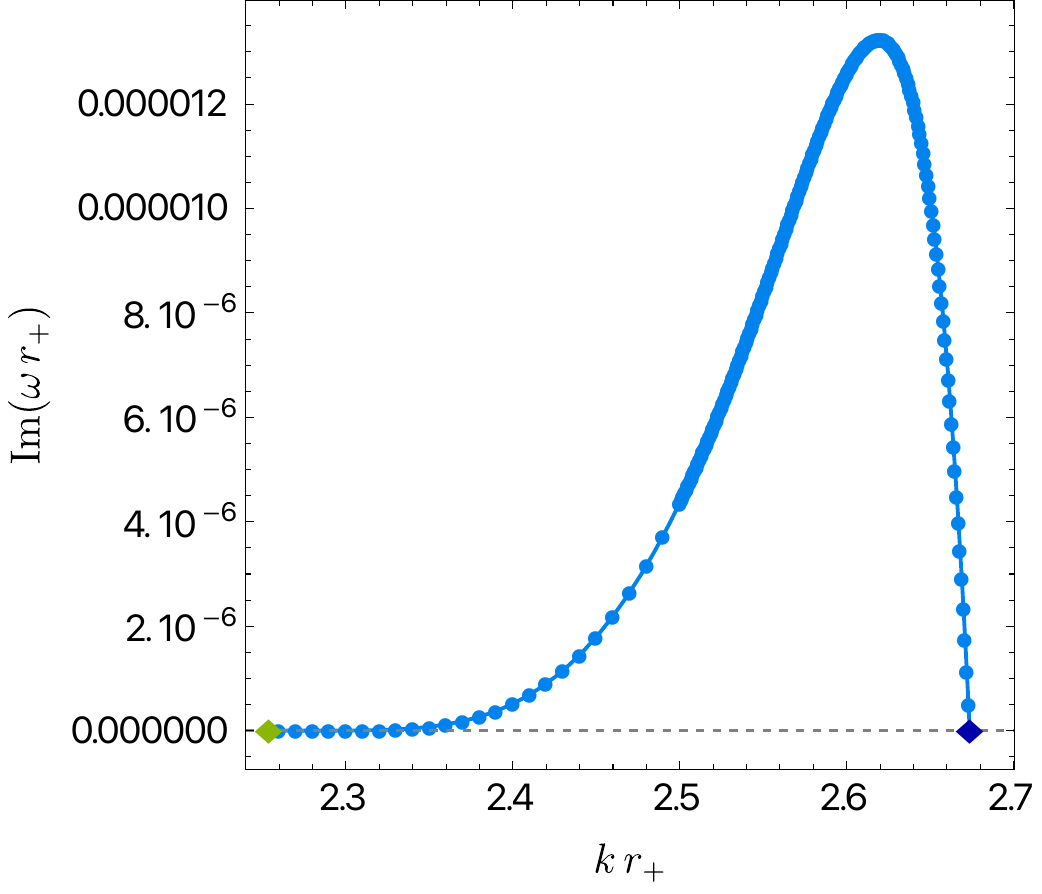}
\hspace{1cm}
\includegraphics[width=.425\textwidth]{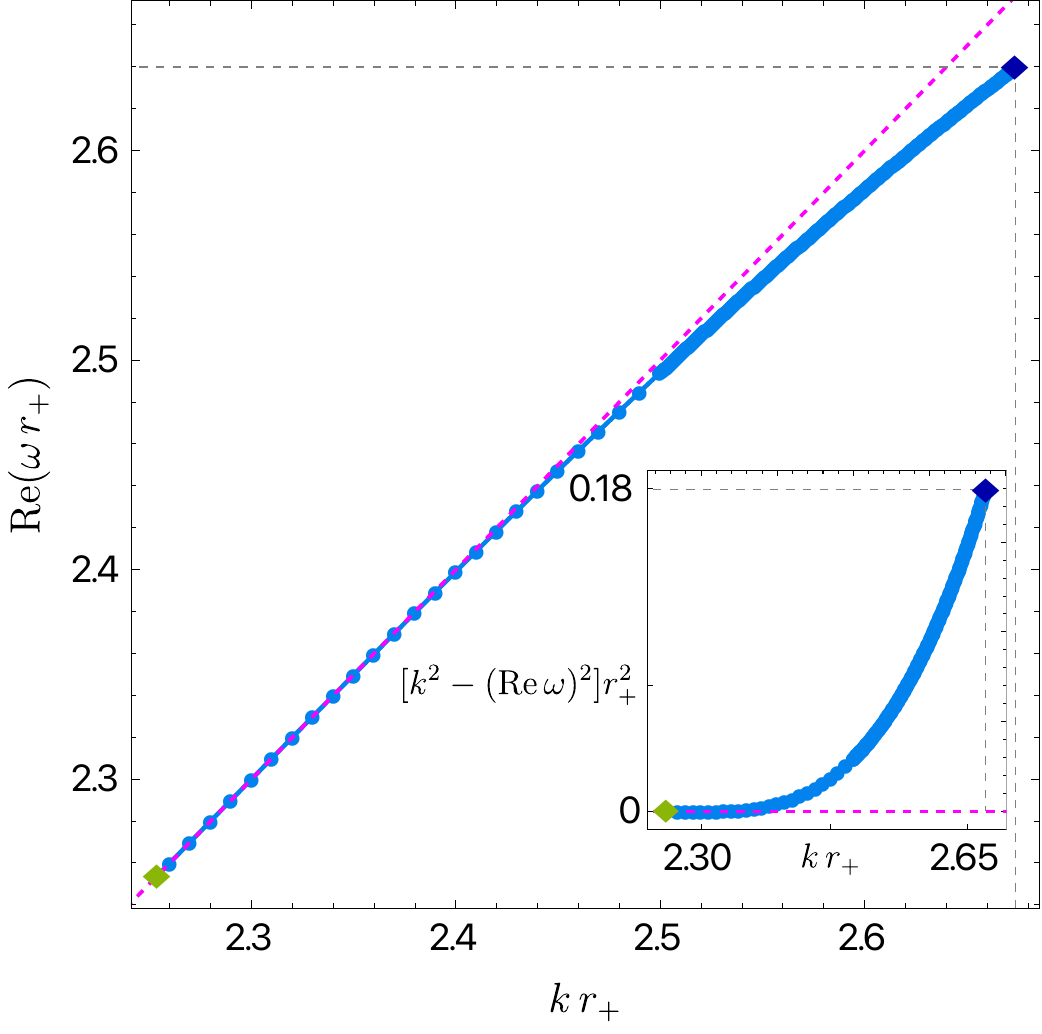}
\caption{Frequency for the unstable superradiant mode of Myers-Perry black strings with fixed $\widetilde{\Omega}_H=0.66$, as a function of wavenumber $kr_+$.  {\bf Left:} The imaginary part of the frequency ${\rm Im}(\omega r_+)$ is positive for $kr_+\in (\widetilde{k}_{\star},\widetilde{k}_{(0)})$ where the system is unstable to superradiance. The blue diamond on the right with $kr_+=\widetilde{k}_{(0)}\simeq 2.673761$ corresponds to the superradiance onset found in Fig.~\ref{Fig:zeroMode}. One has ${\rm Im}(\omega r_+)\to 0$ as $k r_+\to \widetilde{k}_{\star}=\frac{3 \sqrt{1411}}{50}$ (green diamond on the left side of the plot).  
{\bf Right:} The magenta dashed line describes the curve Re$(\omega r_+)=k r_+$. We see that ${\rm Re}(\omega r_+)<k r_+$, and that ${\rm Re}(\omega r_+)\to kr_+$ as $k r_+\to\widetilde{k}_{\star}$ from above (this is best seen in the inset plot which displays $\widetilde{k}^2-(\mathrm{Re}\, \widetilde{\omega})^2$ as a function of $\widetilde{k}$).  This demonstrates that when the instability shuts down at $kr_+=\widetilde{k}_{\star}$, the fall-off $e^{-\sqrt{k^2-\omega^2}\,r}$ ceases to be exponential.
}
\label{Fig:timescale}
\end{figure}

To gather more evidence, we focus our attention on Myers-Perry black strings with a fixed $\widetilde\Omega_H$ and compute the frequency $\widetilde\omega$ while scanning over $\widetilde k$. The results for $\widetilde{\Omega}_H=0.66$ are shown in Fig.~\ref{Fig:timescale}.  The left panel demonstrates that the instability is shutting off as $\widetilde{k}$ approaches $\widetilde{k}_{\star}(0.66)=\frac{3 \sqrt{1411}}{50}\simeq 2.253797$ from above (which is the point in the curve $cB$ of Fig.~\ref{Fig:zeroMode} with $\widetilde{\Omega}_H=0.66$). The right panel demonstrates that as this occurs, the exponential falloff $e^{-\sqrt{k^2-\omega^2}\,r}$ is becoming marginal.

\begin{figure}[t]
\centering
\includegraphics[width=.45\textwidth]{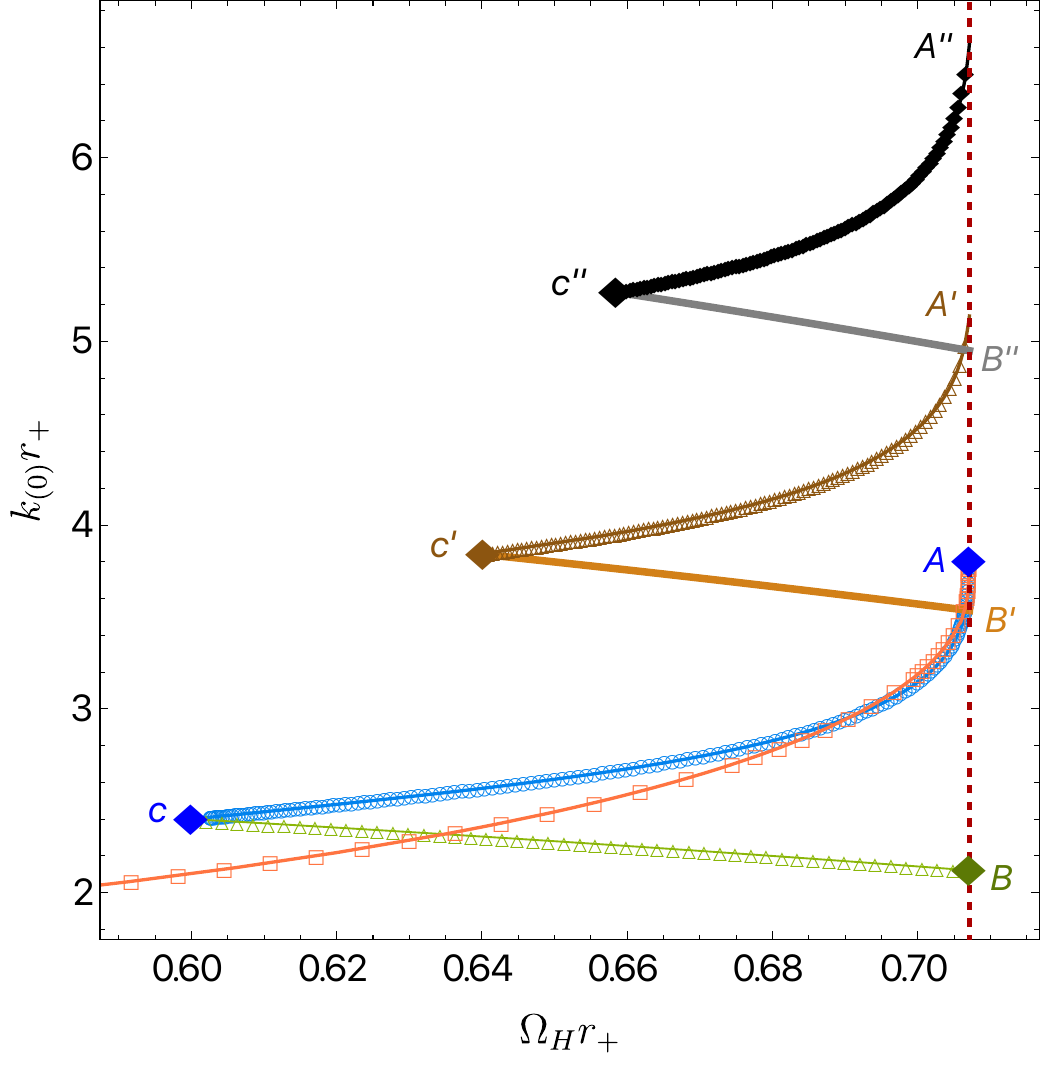}
\hspace{1cm}
\includegraphics[width=.45\textwidth]{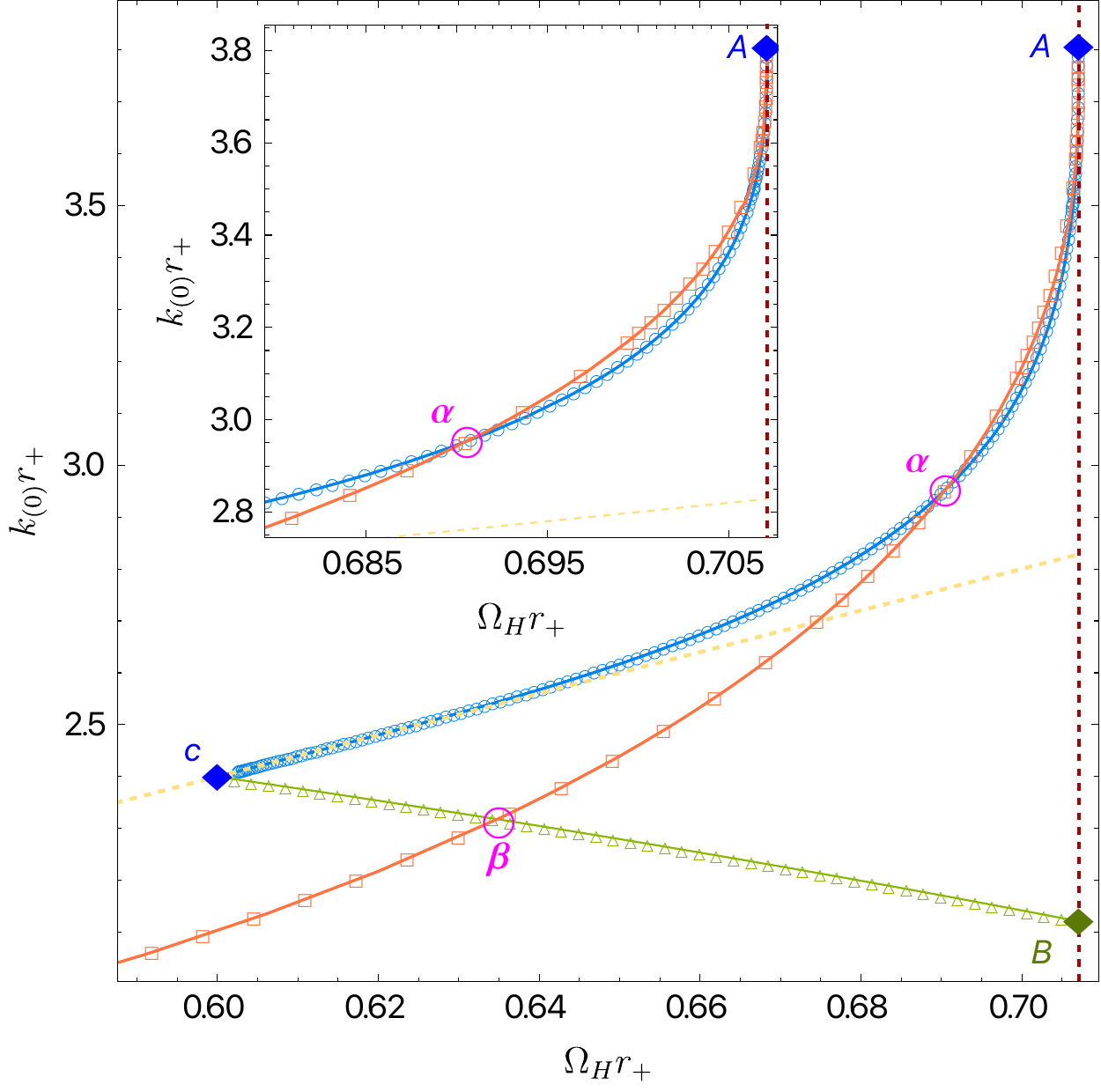}
\caption{
 {\bf Left panel:} Regions $ABc$, $A'B'c'$, and $A''B''c''$ where Myers-Perry black strings are unstable to $m=2,3,4$ superradiant modes, respectively. The orange squares correspond to the onset of the Gregory-Laflamme instability already shown in Figs.~\ref{Fig:GLmodes} and \ref{Fig:zeroMode} (the region below this curve is unstable).
{\bf Right panel:} 
Zoom-in of region unstable to both Gregory-Laflamme and superradiance. $\alpha$ and $\beta$ mark where the Gregory-Laflamme onset curve intersects with the $m=2$ unstable superradiant region. $\alpha$ lies at $(\widetilde{\Omega}_H, \widetilde{k}_{(0)}, \widetilde{L}_{(0)})_{\alpha}\simeq (0.69054,2.95350,2.12737)$, where to the left of $\alpha$ one has $\widetilde{k}_{(0)}\big|_{\hbox{\tiny SR}}
>\widetilde{k}_{(0)}\big|_{\hbox{\tiny GL}}$, while to the right of $\alpha$ one has $\widetilde{k}_{(0)}\big|_{\hbox{\tiny SR}}<\widetilde{k}_{(0)}\big|_{\hbox{\tiny GL}}$.
$\beta$ lies at $(\widetilde{\Omega}_H, \widetilde{k}_{(0)}, \widetilde{L}_{(0)})_{\beta}\simeq (0.63478,2.31810,2.71049)$.
} \label{Fig:m3m4}
\end{figure}

With this evidence at hand, we can now attempt to find the green curve (hereafter referred to as the confining cutoff curve) $\widetilde{k}_{\star}(\widetilde{\Omega}_H)$ directly by enforcing $\widetilde\omega=\widetilde k$ in our ODE.  This requires that we revisit the asymptotic Frobenius analysis of \eqref{EOMsuper}. Indeed, from \eqref{PertVariable}, we know that generic superradiant modes decay as $e^{-\frac{\sqrt{\widetilde{k}^2-\widetilde{\omega}^2}}{1-y^2}}$. However, when ${\rm Re}\,\widetilde{\omega}=\widetilde{k}$ and ${\rm Im}\,\widetilde{\omega}=0$, the argument of this exponential vanishes. For generic $m$, a Frobenius analysis now gives the asymptotic decays $Q\sim (1-y)^{1\pm\gamma}$ with
\begin{equation}\label{gammaGrav}
 \gamma = \sqrt{(2m-1)^2 - \frac{\widetilde{k}^2}{1-\widetilde{\Omega}_H^2}}\ .
\end{equation}
It is not clear which combination of these modes we should take as a boundary condition. However, note that there is a further special case: if $\gamma=0$, then the two decays above degenerate to $Q\sim (1-y)$ and the second independent solution now decays logarithmically.  For reasons that are not completely understood, the curve $\widetilde{k}^{(m)}_{\star}(\widetilde{\Omega}_H)$ satisfies this degenerate case.
That is to say, apparently, $\gamma=0$ is the borderline behaviour for whether the asymptotic behaviour is oscillatory or not and turns out to be the condition for $\widetilde{k}^{(m)}_{\star}(\widetilde{\Omega}_H)$.  We have verified this empirically by directly solving for such modes, and by verifying that such modes are approached as $\widetilde{k}\to \widetilde{k}^{(m)}_{\star}$ (see again Fig.~\ref{Fig:timescale}).  From this evidence, and solving for $\widetilde k$ from the condition $\gamma=0$, we conclude that
\begin{equation}\label{CutoffHighm}
\widetilde{k}^{(m)}_{\star} =(2m-1)\sqrt{1-\widetilde{\Omega}_H^2}\;,
\end{equation}
which is an explicit analytic expression for $\widetilde k^{(m)}_{\star}(\widetilde\Omega_H)$ that is fully consistent with our numerical results. For $m=2$ this gives the cutoff curve $cB$ of Fig.~\ref{Fig:zeroMode} (and for $m=2,3,4$ this gives the curves $cB$, $c'B'$ and $c''B''$, respectively, of later Fig.~\ref{Fig:m3m4}). Note, though that the analytic expression \eqref{CutoffHighm} is based on an assumption on the asymptotic behaviour of the linear solution for which we can only justify empirically.  Furthermore, obtaining this result does not actually require us to solve a full boundary value problem, for which the boundary conditions remain unclear.  It remains a possibility that the actual linear solution in the limit of the cutoff curve might not be regular.  Indeed, to obtain \eqref{CutoffHighm} we did not have to require, as a boundary condition, that the logarithmic solution must vanish.

Now we can also determine the frequency $\widetilde{\Omega}^{(m)}_H$ where $\widetilde{k}^{(m)}_{(0)}$ and $\widetilde{k}^{(m)}_{\star}$ intersect; for $m=2$ this is point $c$ in Fig.~\ref{Fig:zeroMode} where the zero mode $cA$ and cutoff $cB$ curves meet. Recall that $\widetilde{k}^{(m)}_{(0)}$ satisfies $\widetilde{\omega}=2m\widetilde{\Omega}_H$ and $\widetilde{k}^{(m)}_{\star}$ satisfies $\widetilde{\omega}=\widetilde{k}$.  Combining these with \eqref{CutoffHighm}, we find that these curves meet at
\begin{equation}\label{OmegaChighm}
\widetilde{\Omega}_H^{(m)}=\frac{(2 m-1)}{\sqrt{8 m^2-4 m+1}}  \quad \Leftrightarrow \quad \frac{{\Omega}_H^{(m)}}{\Omega^{\mathrm{ext}}_H}=\frac{\sqrt 2 (2 m-1)}{\sqrt{8 m^2-4 m+1}} \,.
\end{equation}

The superradiant onset in the extremal Myers-Perry limit can be also obtained analytically (for $m=2$, this is point $A$ in Fig.~\ref{Fig:zeroMode}).
In the extremal Myers-Perry limit $\widetilde{\Omega}_H=\widetilde{\Omega}^{\mathrm{ext}}_H=1/\sqrt{2}$ and at the onset of the superradiant instability $\widetilde{\omega}=2m\widetilde{\Omega}_H$, the perturbation equation~(\ref{EOMsuper}) becomes
\begin{equation}
 Q''+\frac{3r^2+r_+^2}{r(r^2-r_+^2)}Q'-\frac{r^2 \{(k^2 r_+^2  - 2 m^2) r^2 - 4 m r_+^2\}}{r_+^2(r^2-r_+^2)^2}Q=0\ .
\end{equation}
At the horizon, $Q$ behaves as $Q=(r-r_+)^\eta$ with $\eta=\left(-1\pm\sqrt{\widetilde{k}^2 - (2 m^2 + 4m -1)}\,\right)/2$.
For reasons that are again not completely understood, the onset of instability at extremality seems empirically to be located at the boundary of the oscillating and non-oscillating behaviors, when we have the degenerate case $\eta=-1$
We therefore have
\begin{equation}\label{OnsetR=Ext}
\widetilde{k}_{(0)}^{(m)} |_{\hbox{\tiny ext}}=\sqrt{2 m^2 + 4m -1}\,,
\end{equation}
which gives an analytic expression for the onset mode wavenumber at extremality.
For $m=2$, we have $\widetilde{k}=\sqrt{15}\simeq 3.873$ which is point $A$ in Fig.~\ref{Fig:zeroMode} where we also see that our numerical results for the onset curve $\widetilde{k}^{(m)}_{(0)}(\widetilde{\Omega}_H)$ approach $A$ as given by \eqref{OnsetR=Ext} in the extremal limit. Moreover, for $m=3$ and $m=4$ equation \eqref{OnsetR=Ext} also gives the points $A'$ and $A''$ in later Fig.~\ref{Fig:m3m4}.

We can now show the superradiant onset curve and confining cutoff curves $\widetilde{k}^{(m)}_{(0)}$ and $\widetilde{k}^{(m)}_{\star}$ also for other values of $m$.  The left panel of Fig.~\ref{Fig:m3m4} show the results for $m=2,3,4$. (We also obtained qualitatively similar curves for half-integer modes $m=5/2,7/2$, but do not show them to avoid clutter.)  The unstable regions $A'B'c'$ and $A''B''c''$ for $m=3$ and $m=4$, respectively are qualitatively similar to the unstable region $ABc$ for $m=2$ but transported to higher values of $\widetilde{k}$ as $m$ grows. The cutoff curves $cB$, $c'B'$ and $c''B''$ are described by \eqref{CutoffHighm} with $m=2,3,4$, respectively, and points $c$, $c'$ and $c''$ are described by \eqref{OmegaChighm} with $m=2,3,4$, respectively. The zero mode curves $cA$, $c'A'$ and $c''A'$ for $m=2,3,4$ were obtained numerically solving the eigenvalue problem for $\widetilde{k}^{(m)}_{(0}(\widetilde{\Omega}_H)$, which approach the points $A$, $A'$ and $A''$ at extremality as given by \eqref{OnsetR=Ext}. Perturbations with adjacent $m$'s also overlap in some small regions of parameter space close to extremality. So there are Myers-Perry strings that are superradiant unstable to a few $m$'s (but not an infinite number of them like in other superradiant black hole systems).

We also see that the unstable regions ($ABc$, $A'B'c'$, $A''B''c''$, ...) decrease as $m$ increases. Essentially, this occurs because the intersection points $c, c', c'',\ldots$ described by \eqref{OmegaChighm} have an angular velocity that is increasingly close to $\widetilde{\Omega}_H^{\rm ext}$ as $m$ increases, being exactly $\widetilde{\Omega}_H^{\rm ext}$ in the limit $m\to\infty$. So in the eikonal limit ($m\to\infty$), the unstable region becomes a measure zero region in parameter space and the eikonal limit of points $A, B$ and $c$ all collapse to a common single point at $\widetilde{\Omega}_H=\widetilde{\Omega}_H^{\rm ext}$ and $\widetilde{k}\to \infty$, as follows from \eqref{OmegaChighm} and \eqref{OnsetR=Ext} in the limit $m\to\infty$.  That is to say, the superradiant instability for Myers-Perry black strings shuts down as $m\to\infty$, which is in agreement with the fact that the Myers-Perry black string has no stable circular geodesics \cite{Tangherlini:1963,Frolov:2003en}. In this precise sense, this is a `{\it finite-m}' superradiant instability that occurs in black strings, but is not observed in (known) superradiant black holes (see also the discussion in footnote~\ref{foot:massiveScalar}).

In Fig.~\ref{Fig:m3m4}, we also include the onset curve (orange squares) for the Gregory-Laflamme instability that we have already computed in section~\ref{sec:GL}.  We see that there are regions of parameter space where either, none, or both types of instabilities are present. 

\begin{figure}[t]
\centering
\includegraphics[width=.535\textwidth]{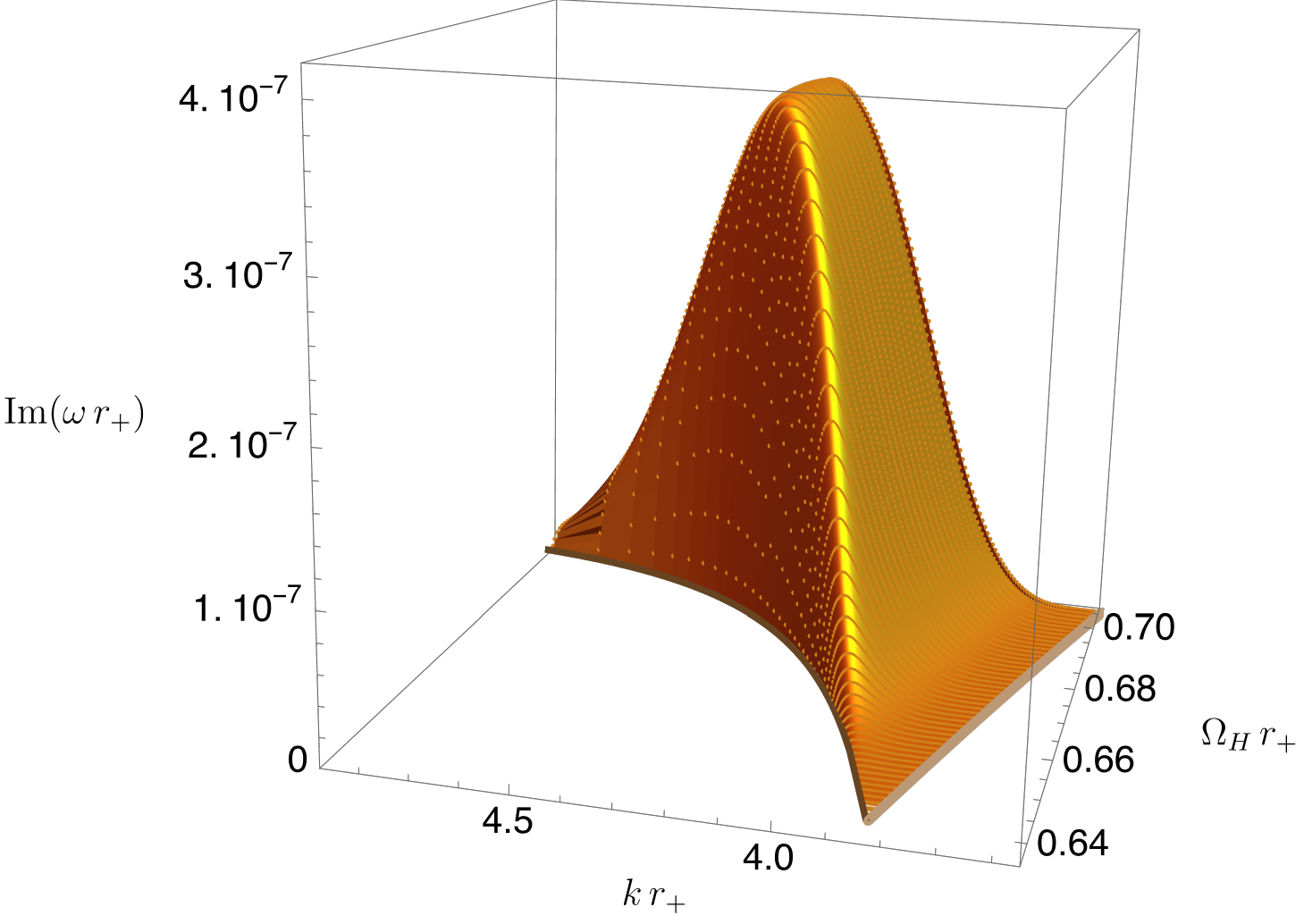}
\hspace{-0.2cm}
\includegraphics[width=.46\textwidth]{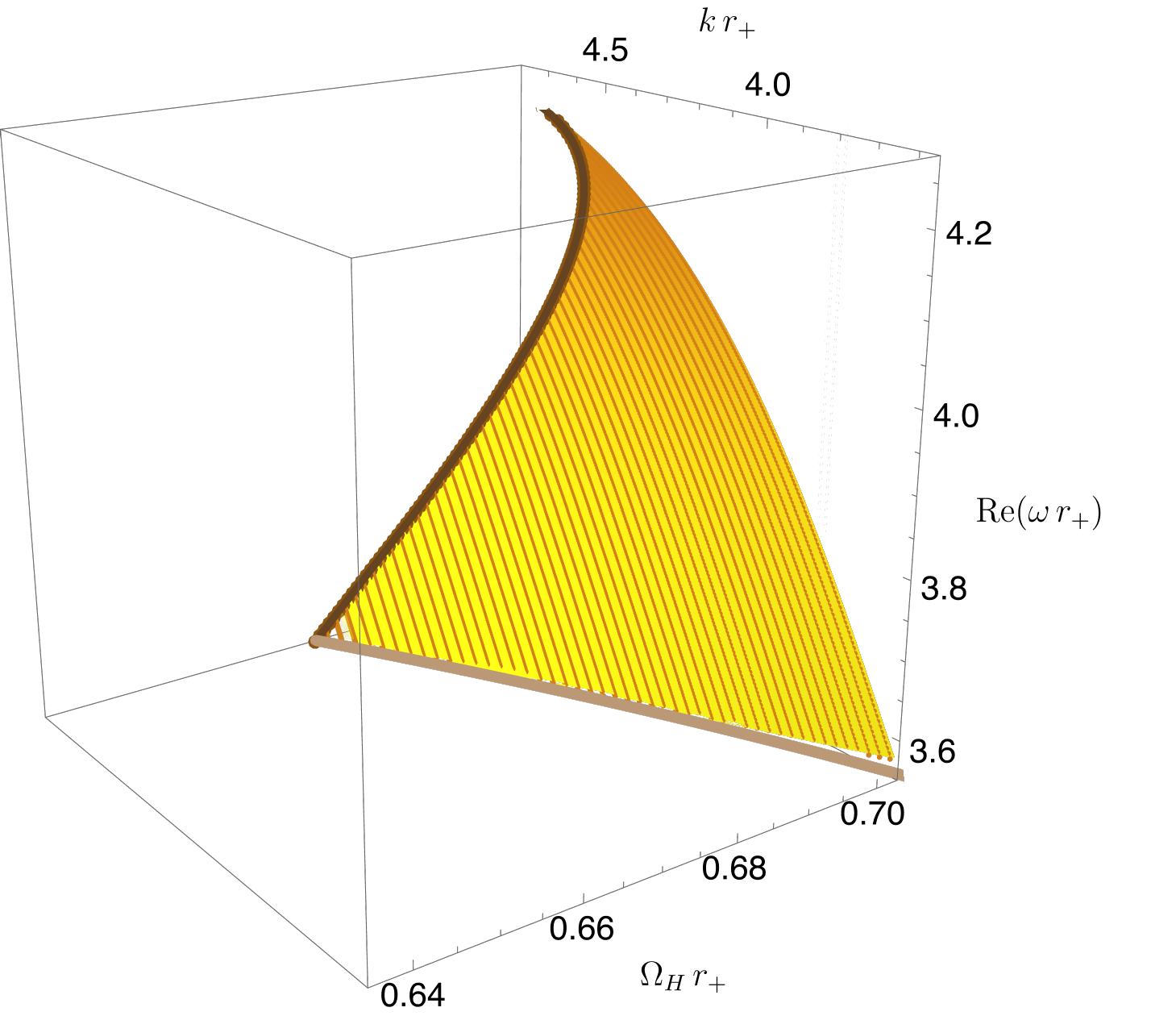}
\caption{Frequency for $m=3$ unstable superradiant modes, as a function of parameters of Myers-Perry black strings. The instability is present (\ie ${\rm Im}\,\widetilde{\omega}>0$) for $  \widetilde{\Omega}_H|_{c'} \leq  \widetilde{\Omega}_H \leq 1/\sqrt{2}$, with $ \widetilde{\Omega}_H|_{c'} =\frac{5}{\sqrt{61}}$, and $  \widetilde{k}_{\star}(\widetilde{\Omega}_H)\leq \widetilde{k}\leq \widetilde{k}_{(0)}(\widetilde{\Omega}_H)$. The zero mode curve $\widetilde{k}_{(0)}(\widetilde{\Omega}_H)$ (and ${\rm Im}\,\widetilde{\omega}=0$) was  computed independently (see $c'A'$) in Fig.~\ref{Fig:m3m4} and we  show it as a continuous brown line in the ${\rm Im}\,\widetilde{\omega}=0$ plane. On the other hand, the light-brown curve represents the confining cutoff line  $\widetilde{k}_{\star}(\widetilde{\Omega}_H)$ as defined parametrically in  \eqref{CutoffHighm} for $m=3$ (i.e. the cutoff curve $c'B'$ in Fig.~\ref{Fig:m3m4}). It meets the zero mode curve at $ \widetilde{\Omega}_H|_{c'}$.
}
\label{Fig:timescaleSuper-m3}
\end{figure}

\begin{figure}[t]
\centering
\includegraphics[width=.505\textwidth]{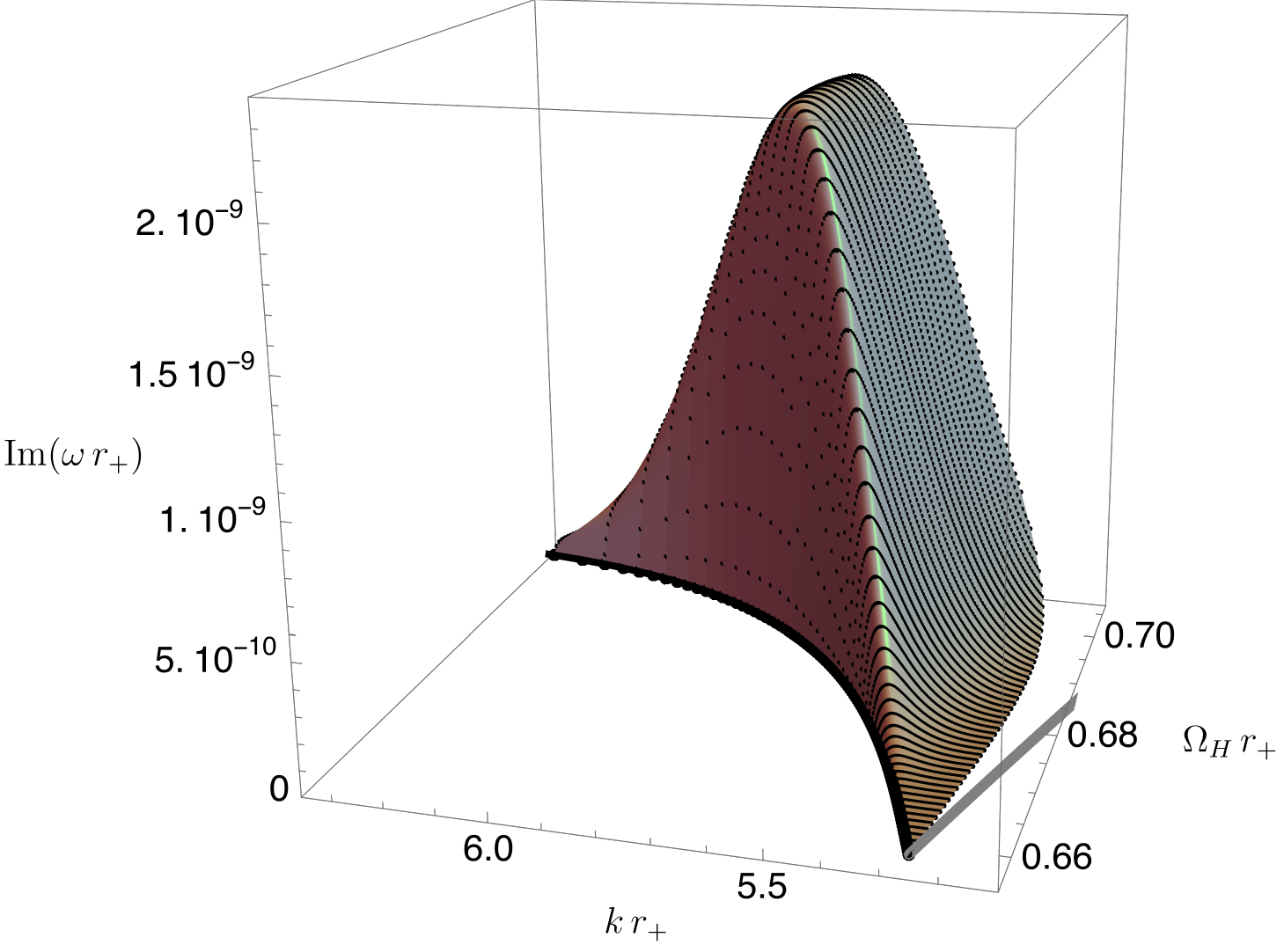}
\hspace{-0.2cm}
\includegraphics[width=.49\textwidth]{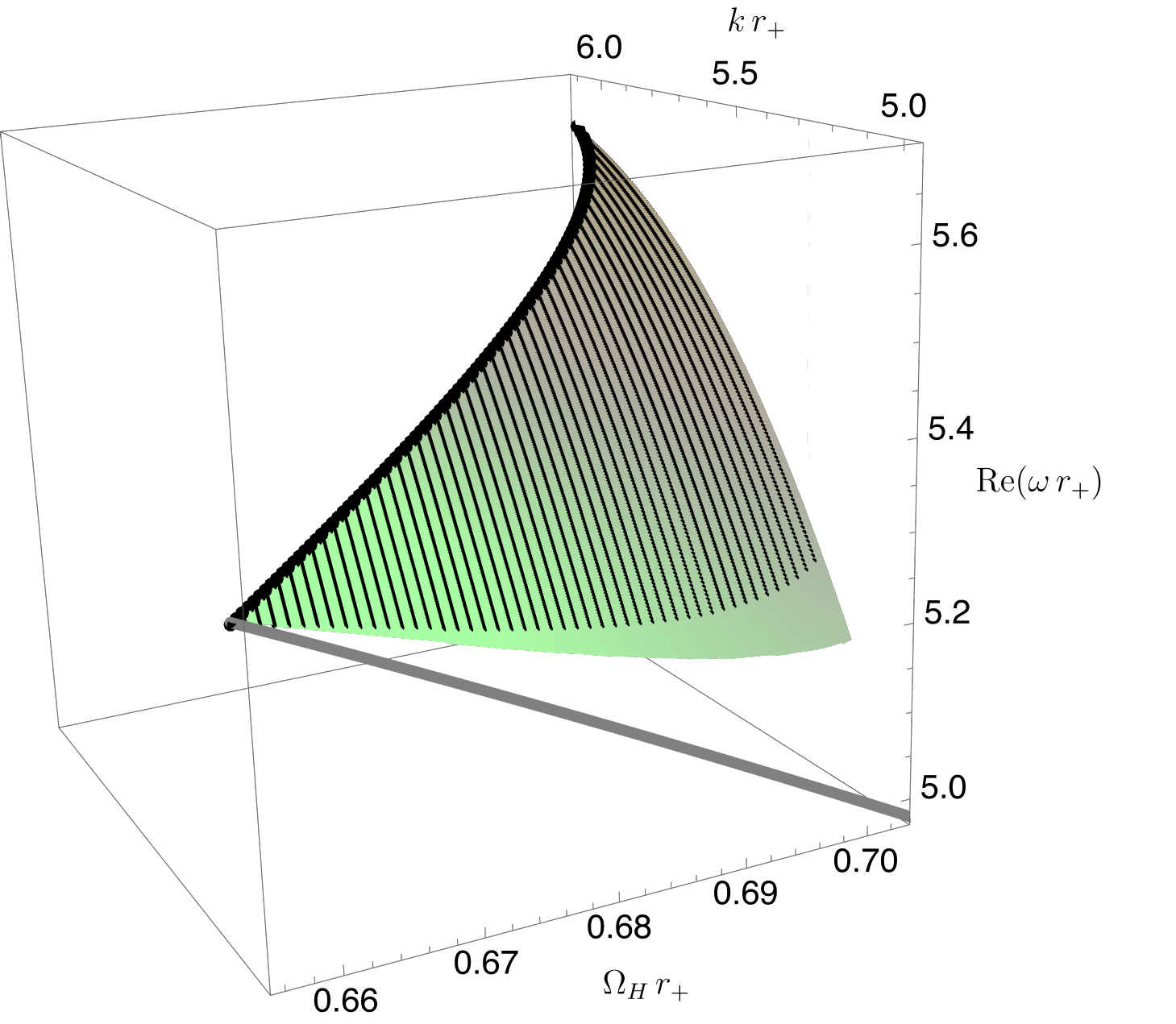}
\caption{Frequency for $m=4$ unstable superradiant modes, as a function of parameters of Myers-Perry black strings.
The instability is present (\ie ${\rm Im}\,\widetilde{\omega}>0$) for $  \widetilde{\Omega}_H|_{c''} \leq  \widetilde{\Omega}_H \leq 1/ \sqrt{2}$, with $ \widetilde{\Omega}_H|_{c''} =\frac{7}{\sqrt{113}}$, and $  \widetilde{k}_{\star}(\widetilde{\Omega}_H)\leq \widetilde{k}\leq \widetilde{k}_{(0)}(\widetilde{\Omega}_H)$. The zero mode curve $\widetilde{k}_{(0)}(\widetilde{\Omega}_H)$ (and ${\rm Im}\,\widetilde{\omega}=0$) was  computed independently (see $c''A''$) in Fig.~\ref{Fig:m3m4} and we  show it here as a continuous black line in the ${\rm Im}\,\widetilde{\omega}=0$ plane. On the other hand, the  gray curve represents the line  $\widetilde{k}_{\star}(\widetilde{\Omega}_H)$ as defined parametrically in  \eqref{CutoffHighm} for $m=4$ (i.e. the cutoff curve $c''B''$ in Fig.~\ref{Fig:m3m4}). It meets the zero mode black curve at $ \widetilde{\Omega}_H|_{c''}$. It is very difficult to obtain solutions in the neighborhood of the cutoff gray line $\widetilde{k}_{\star}(\widetilde{\Omega}_H)$ so we do not attempt to find them in the present $m=4$ case; hence we find a gap between when the surface approaches this gray borderline.
}
\label{Fig:timescaleSuper-m4}
\end{figure}

For completeness, in Figs.~\ref{Fig:timescaleSuper-m3} and \ref{Fig:timescaleSuper-m4}, we show the real and imaginary parts of the frequency for $m=3$ and $m=4$ superradiant instabilities (in the same manner as the $m=2$ case in Fig.~\ref{Fig:timescaleSuper}).  We find that for each $m$, increasing the rotation also (typically) increases the growth rate.  We also see that increasing $m$ by one roughly reduces the growth rate by two orders of magnitude.

Until now, we have parametrised the Myers-Perry black strings using $\widetilde\Omega_H$ and $\widetilde k$.  This was for the convenience of computation and presentation. However, for the purposes of discussing the time evolution of the instabilities, as well as the associated novel black string solutions that branch from the onset of these instabilities, it is more convenient to use the conserved quantities of energy and angular momenta as parameters.  In time evolution, it is natural to keep the Kaluza-Klein circle fixed (i.e. we fix its length $L=\widetilde{L} r_+$), so it is natural to express quantities in units of $L$.

From  \eqref{ThermoMP}, the dimensionless energy,  $\mathcal{E}\equiv E/L^3$, and dimensionless angular momenta,  $\mathcal{J}\equiv J/L^4$, are given by
\begin{equation}\label{ThermoMPonset}
\mathcal{E}=\frac{1}{G_6}\,\frac{3 \pi}{8\widetilde{L}^2} \frac{1}{1-\widetilde{a} ^2}=\frac{1}{G_6}\,\frac{3\widetilde{k}^2}{32\pi(1-\widetilde{\Omega}_H^2)}\,, \qquad \mathcal{J}=\frac{1}{G_6}\,\frac{\pi}{4 \widetilde{L}^3}\, \frac{\widetilde{a}}{1-\widetilde{a} ^2}=\frac{1}{G_6}\,\frac{\widetilde{k}^3\widetilde{\Omega}_H}{32\pi^2(1-\widetilde{\Omega}_H^2)}\,.
\end{equation}
At extremality, we have
\begin{equation}\label{MPext:EJ}
\mathcal{J}^{\hbox{\tiny ext}}(\mathcal{E})=\frac{2^{3/2}}{3^{3/2}\pi^{1/2}} G_6^{1/2}\mathcal{E}^{3/2}\;.
\end{equation}

\begin{figure}[t]
\centerline{
\includegraphics[width=.55\textwidth]{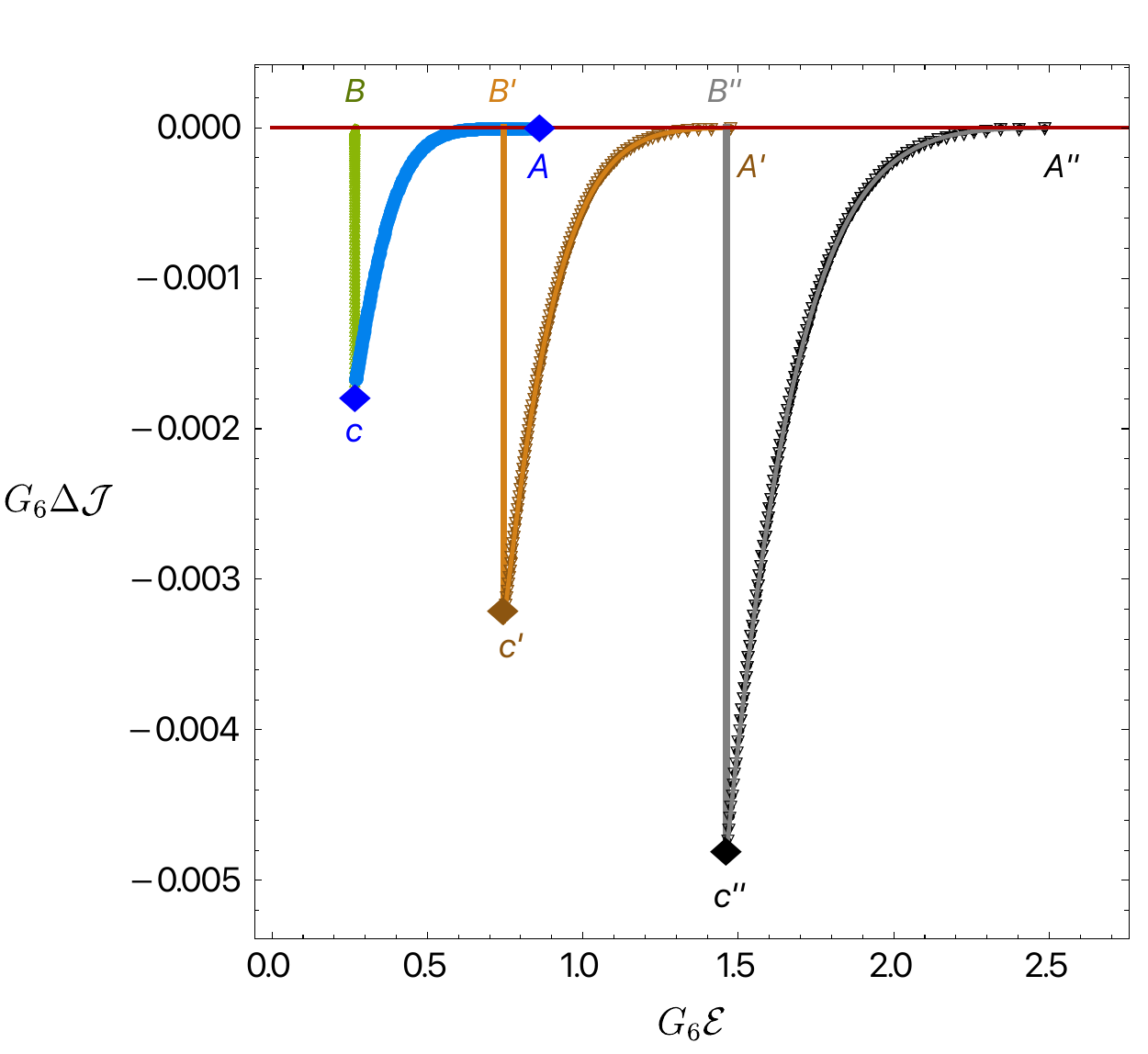}
\hspace{0.1cm}
\includegraphics[width=.55\textwidth]{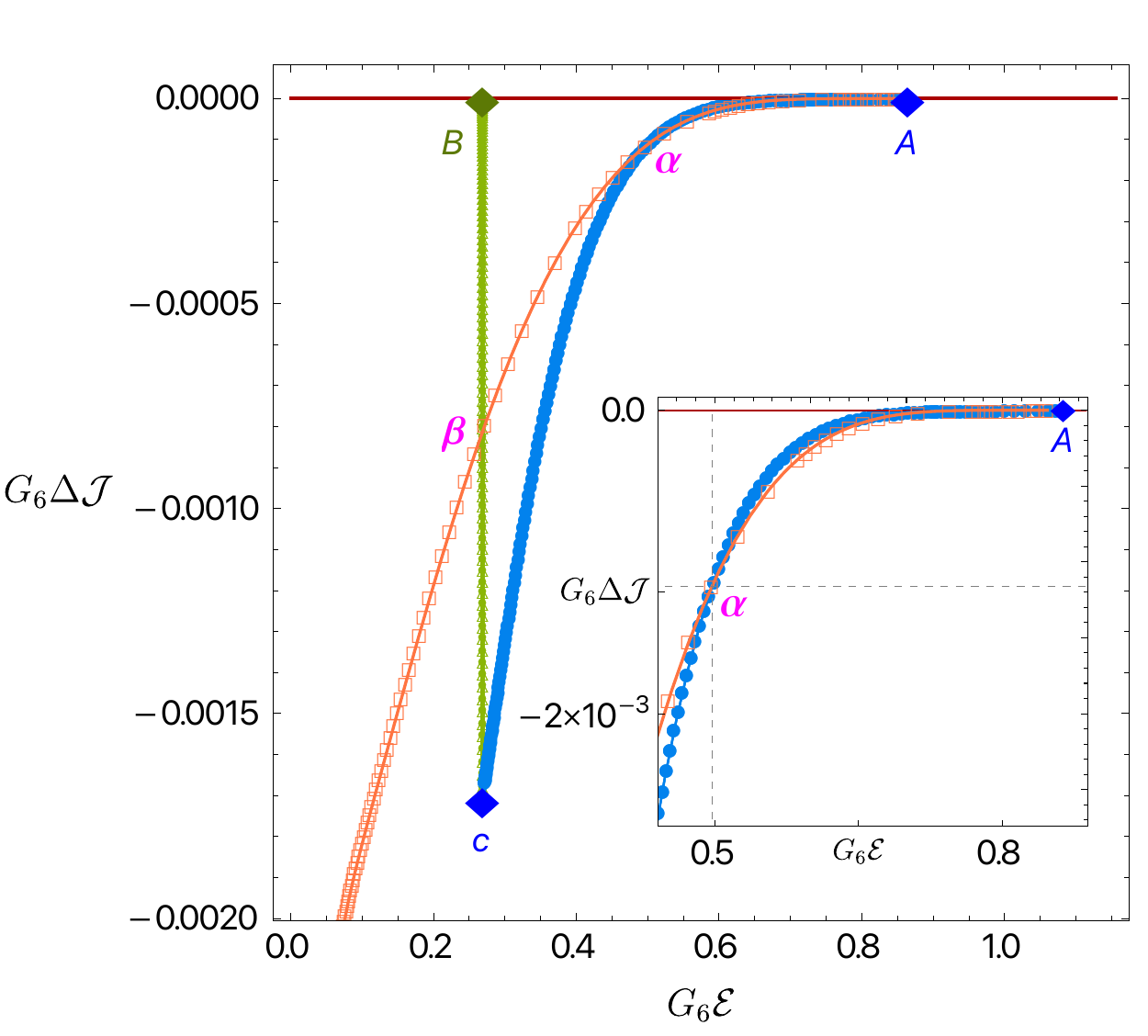}
}
\caption{Onset/cutoff curves for the $m=2,3,4$ superradiant instability and the zero mode curve of Gregory-Laflamme instability (orange squares on right panel). The vertical axis describes the difference between a given solution and the extremal Myers-Perry solution that has the same energy, $\Delta \mathcal J\equiv \left(\mathcal J-\mathcal J^{\mathrm{ext}}\right)|_{\hbox{\tiny same}\, \mathcal E}$.
Equally-spinning Myers-Perry black strings are unstable to $m=2$, $m=3$ and $m=4$ superradiant perturbations if they lie inside the regions $ABc$, $A'B'c'$ and $A''B''c''$, respectively. They are unstable to the $m=0$ Gregory-Laflamme instability if they are contained above the orange square curve (and below the horizontal extremal curve with $\Delta\mathcal{J}=0$) in the right panel. All the points in this figure are in a 1-to-1 correspondence to the same points in Fig.~\ref{Fig:m3m4}.
}
\label{fig:stabilityDiag}
\end{figure}

We can now use \eqref{ThermoMPonset} to translate the onset and cutoff curves of  Figs.~\ref{Fig:zeroMode}~and~\ref{Fig:m3m4} into the $\mathcal{E}$-$\mathcal{J}$ phase diagram.
For example, the $m=2$ superradiant onset $cA$ curve $\widetilde{k}_{(0)}(\widetilde{\Omega}_H)$ of Fig.~\ref{Fig:zeroMode}, for $\widetilde{\Omega}_H|_c \leq \widetilde{\Omega}_H \leq 1/\sqrt{2}$, defines via \eqref{ThermoMPonset}  the onset curve for the $m=2$ superradiant instability of Myers-Perry strings in the  $\mathcal{E}$-$\mathcal{J}$ phase diagram. This curve turns out to be very close to the curve that describes extremal Myers-Perry strings described by \eqref{MPext:EJ}. Therefore, to have a plot where the various unstable regions can be presented clearly, it is convenient to plot instead $\mathcal{E}$ {vs} $\Delta \mathcal{J}\equiv (\mathcal{J}-\mathcal{J}^{\hbox{\tiny ext}})|_{\hbox{\tiny same}\,\mathcal{E}}$ where the latter describes the angular momentum difference between a given solution $-$ \eg the superradiant onset or cutoff Myers-Perry family $-$ and the extremal Myers-Perry string at the same $\mathcal{E}$. This phase diagram $\mathcal{E}$-$\Delta\mathcal{J}$ is displayed in Fig.~\ref{fig:stabilityDiag}.

In Fig.~\ref{fig:stabilityDiag}, Myers-Perry black strings exist with $\mathcal{E}>0$ for arbitrarily large $\mathcal{E}$, and for any angular momenta at or below extremality $\Delta \mathcal{J}\leq 0$.  The triangular region $ABc$ (respectively, $A'B'c'$ and $A''B''c''$) is unstable to $m=2$ ($m=3$ and $m=4$) superradiant modes.\footnote{The horizontal red line with $\Delta \mathcal{J}=0$ represents the 1-parameter family of extremal Myers-Perry strings with $\widetilde{\Omega}_H=1/\sqrt{2}$. It extends to arbitrarily large $\mathcal{E}$. Non-extremal Myers-Perry strings exist below this line. On the other hand, the blue disk curve $Ac$ represents Myers-Perry strings at the onset of the $m=2$ superradiant instability, as described by \eqref{ThermoMPonset} with $\widetilde{k}\equiv \widetilde{k}_{(0)}(\widetilde{\Omega}_H)$, for $\widetilde{\Omega}_H|_c \leq \widetilde{\Omega}_H \leq 1/\sqrt{2}$. The endpoints of this zero mode $cA$ curve are determined by the solutions  \eqref{ThermoMPonset} with $(\widetilde{\Omega}_H, \widetilde{k}_{(0)})|_c= (3/5,12/5)$  and  $(\widetilde{\Omega}_H, \widetilde{k}_{(0)})|_A=(1/\sqrt{2}, \sqrt{15})$, as follows from \eqref{OmegaChighm} and \eqref{OnsetR=Ext}. The confining cutoff curve for $m=2$ superradiant family, namely the curve $cB$ in  Fig.~\ref{Fig:zeroMode} or \ref{Fig:m3m4} is described by $\widetilde{k}^{(m)}_{\star}(\widetilde{\Omega}_H)$ in \eqref{CutoffHighm}. This translates, after using \eqref{ThermoMPonset}, into a curve with constant $G_6 \mathcal{E}=27/(32\pi)$ and $G_6 \mathcal{J}=27\widetilde{\Omega}_H\sqrt{1-\widetilde{\Omega}_H^2}/(32\pi^2)$ with  $3/5\leq \widetilde{\Omega}_H \leq 1/\sqrt{2}$ in the  $\mathcal{E}$-$\mathcal{J}$ phase diagram. This gives the vertical line $Bc$ in the phase diagram $\mathcal{E}$ {vs} $\Delta \mathcal{J}$ of Fig.~\ref{fig:stabilityDiag}. We can now repeat this exercise for the Gregory-Laflamme and $m=3,4$ superradiant curves of Fig.~\ref{Fig:m3m4} to find the associated onset/cutoff curves for these modes in  the $\mathcal{E}$ {vs} $\Delta \mathcal{J}$ phase diagram of Fig.~\ref{fig:stabilityDiag}.}
The region above the orange squares line $A\alpha\beta\ldots$ (and below the red horizontal extremal line) in the right panel of Fig.~\ref{fig:stabilityDiag} are Gregory-Laflamme unstable.

Finally, recall that although we have obtained qualitatively similar curves for $m=5/2,7/2$ but do not show them to avoid clutter. Also, recall from the discussions of section~\ref{sec:GL1} that while there are Gregory-Laflamme type instabilities with angular dependence \cite{Dias:2010eu}, they only exist above a critical rotation parameter and typically have lower growth rate.  Additionally, Myers-Perry black strings that are unstable to these other Gregory-Laflamme modes, are also unstable to the one we consider in Fig.~\ref{fig:stabilityDiag}).

\subsection{The many faces of the superradiant onset}\label{sec:Superradiance3}

The superradiant perturbation $\delta \mathrm ds^2_{\hbox{\tiny SR}}$ in \eqref{SRpert} breaks some of the isometries of the Myers-Perry black string, which recall from \eqref{lkv} is $\mathbb{R}_t \times U(1)_z \times U(1)_\psi  \times SU(2)$, where the direction $z$ along the string is assumed to be Kaluza-Klein compactified. In \eqref{SRpert} one can see that $SU(2)$ is manifestly broken if we consider a perturbation with $m > 2$.  However, for $m=2$, we have $\omega=2m\Omega_H=4\Omega_H$ at the onset of the superradiant instability and, in these conditions, the gravitational perturbation \eqref{SRpert} can be written as
\begin{align}\label{SRpert_m2}
h_{MN}\mathrm dx^{M}\mathrm dx^{N}=r^2 e^{-4\,{\rm i}\, \Omega_H t + {\rm i} \, k \, z} \,Q(r) \sigma_{-}^2\ .
\end{align}
This is clearly invariant under $SU(2)$ and hence, when we extend the analysis nonlinearly, one expects that the $m=2$ superradiant onset may be a bifurcation line for a new family of black string solutions that preserve $SU(2)$. However, it is less obvious what other isometries are preserved or broken by the perturbation \eqref{SRpert_m2} and this deserves further discussion. In fact, some linear combinations of the isometries of the original background can be preserved, as we discuss next.

To find  which isometries are broken by the perturbation \eqref{SRpert_m2}, it is convenient to change to a coordinate frame $\{T,Z,\Psi\}$ such that\footnote{In this frame $\{T,Z,\Psi\}$, the background Myers-Perry string is rotating at infinity, and the angular velocity at the horizon vanishes.}
\begin{equation}
T=t\,, \quad Z=z\,, \quad \Psi = \psi -  \Omega_H t + \frac{k}{4} z\,.
\label{rotcoord}
\end{equation}
Their dual vectors are
\begin{equation}
\partial_T = \partial_t + \Omega_H \partial_\psi\,, \quad \partial_Z = \partial_z - \frac{k}{4} \partial_\psi\,, \quad \partial_\Psi = \partial_\psi\,.
\end{equation}
In these coordinates, the isometry group of the background Myers-Perry black string is $\mathbb{R}_T \times U(1)_Z \times U(1)_\Psi \times SU(2)$. But now we explicitly see that, using also the RHS of \eqref{sigmapm}, the $t$- and $z$- dependence in \eqref{SRpert_m2} is totally absorbed in $\Psi$ by the coordinate transformation. That is to say, let $\Sigma_i$ denote the invariant 1-forms for $(\theta,\phi,\Psi)$, i.e. with ~$\psi$ in $\sigma_i$ of \eqref{Eulerforms} replaced by $\Psi$. Then, after the coordinate transformation \eqref{rotcoord}, one finds that the $m=2$ onset superradiant perturbation \eqref{SRpert_m2} simply reads
\begin{equation}\label{SRpert_m2-B}
 h_{MN}\mathrm dx^{M}\mathrm dx^{N}=r^2 \,Q(r) \Sigma_{-}^2\, ,
\end{equation}
where $\Sigma_\pm\equiv (\Sigma_1\mp {\rm i}\,\Sigma_2)/2$. This breaks $U(1)_\Psi$ because  $\Sigma_{-}^2$ depends explicitly on $\Psi$. Including also the complex conjugate contribution to make the perturbation real, the $m=2$ perturbation \eqref{SRpert_m2-B} at the onset of instability finally reads
\begin{equation}\label{SRpertOnset}
h_{MN}\mathrm dx^{M}\mathrm dx^{N} = \frac{r^2}{2} Q(r) \Sigma_{-}^2 + \textrm{c.c.} = \frac{r^2}{4}Q(r)\left(\Sigma_1^2-\Sigma_2^2\right)\,,
\end{equation}
which preserves all the original Myers-Perry symmetries except axisymmetry. This suggests that the onset of the $m=2$ instability might be a branching line to a new family of  nonlinear solutions with $\mathbb{R}_T \times SU(2) \times U(1)_Z$ symmetry. The null generator of the horizon of such solution would be given by $K=\partial_T=\partial_t + \Omega_H \partial_\psi$, which corresponds to the ``helical'' Killing vector generating $U(1)_T$.

There is however another important observation. Indeed, note that the perturbation equation~\eqref{EOMsuper} is invariant under $k\to -k$. Thus, left and right mover modes along the string are absolutely equivalent and thus the onset \eqref{SRpert_m2} of the $m=2$ superradiant instability can be equivalently written as
\begin{equation}\label{SRpert_m2-C}
 h_{MN}\mathrm dx^{M}\mathrm dx^{N}=r^2 e^{-4\,{\rm i}\, \Omega_H t} \left(c_1 \, e^{{\rm i} \, k \, z} + c_2 \, e^{-{\rm i} \, k \, z} \right) \,Q(r) \sigma_{-}^2\, ,
\end{equation}
where $c_1$ and $c_2$ are arbitrary constants.
We can assume $c_1,c_2\in \mathbb{R}$ without loss of generality by using a symmetry that allows for constant shifts of $t$ and $z$.

The interesting observation is that by applying the coordinate transformation \eqref{rotcoord}, one concludes that \eqref{SRpert_m2-C} breaks $U(1)_Z$ unless $c_1=0$ or $c_2=0$. This means that the $m=2$ superradiant onset $Ac$, that we identify in Figs.~\ref{Fig:zeroMode}~and~\ref{fig:stabilityDiag}, describes the onset of two distinct sectors of perturbations:
one preserves $U(1)_Z$ while the other breaks $U(1)_Z$.

It follows that each point of the onset line $Ac$ that we found in Figs.~\ref{Fig:zeroMode}~and~\ref{fig:stabilityDiag} should be, when we extend the analysis beyond linear order, a degenerate bifurcation line to a two-parameter family of new black string solutions. The two parameters correspond to $c_1$ and $c_2$ in \eqref{SRpert_m2-C} at linear order. One special case of this family is when $c_1=0$ or $c_2=0$.  The resulting one-parameter family preserves $U(1)_Z$. As this $U(1)_Z$ describes a spatially helical symmetry that mixes $\partial_z$ and $\partial_\psi$, we call the $U(1)_Z$-preserving solutions \textit{helical black strings}.\footnote{Existence of black strings with helical symmetry has been first found in~\cite{Emparan:2009vd} using the blackfold effective worldvolume theory.}  All other cases break the $U(1)_Z$, while still preserving the $\mathbb{R}_T \times SU(2)$ isometry group.  As it preserves a $U(1)_T$ temporal helical symmetry that mixes $\partial_t$ and $\partial_\psi$, but breaks other spatial symmetries, we call these \textit{black resonator string} family by analogy to similar black hole solutions (without extended directions) \cite{Dias:2011at,Dias:2011tj,Dias:2015rxy,Ishii:2018oms,Ishii:2020muv,Ishii:2021xmn}.

Of course these novel nonlinear solutions can exist because at the onset of the instability, and only here, the mode perturbation is regular not only on the future horizon but also on the past horizon; hence there is room for the existence of a novel family of rotating black strings that are regular simultaneously at the future and past horizons, and that bifurcate from the Myers-Perry black string at the onset of the $m=2$ superradiant instability.  To confirm the existence of these two new families of black strings, we must solve the Einstein equations nonlinearly, \ie beyond the linear analysis done in the present paper. We will do so and find the \textit{black resonator strings} and the \textit{helical black strings} in the future companion papers \cite{Dias:2022str,Dias:2023nbj}, respectively.

 Finally, let us comment on the prospect of additional solutions that branch from the confining cutoff curve $\widetilde{k}^{(m)}_{\star}$.  Here, we indeed have an onset in the sense that $\mathrm{Im}\,\omega=0$, but the boundary conditions for the linear problem on this curve remain unclear, and it is possible that they are not regular (recall the discussion about the logarithmic behaviour below \eqref{CutoffHighm}). We therefore do not seek solutions that branch from this curve.

\section{Discussion  \label{sec:Discussion}}

We have analysed gravitational instabilities on the six-dimensional equal-spinning Myers-Perry black string, \ie the cohomogeneity-1 black strings with equal angular momenta along the two rotation planes of asymptotically Kaluza-Klein spacetimes ($\mathcal{M}^{1,4}\times S^1$). We find instances of both the Gregory-Laflamme instability and the superradiant instability, determined their unstable regions in the 2-dimensional Myers-Perry string parameter space, and computed their growth rates.

Our results on the Gregory-Laflamme instability complement the previous results in \cite{Kleihaus:2007dg,Dias:2010eu} (see also \cite{Dias:2009iu,Dias:2010maa} and \cite{Dias:2011jg}) and supports the broad idea that rotation typically enhances the Gregory-Laflamme instability.

Our results on the superradiant instability are new.  As we have argued, it would be difficult to anticipate the existence of this instability within this system without doing the actual computation.  On one hand, Myers-Perry black strings certainly have an ergoregion where superradiant amplification can occur, and nontrivial superradiant Kaluza-Klein modes are effectively massive and, in some cases, can be confined as first proposed in \cite{Marolf:2004fya} (see also \cite{Cardoso:2004zz,Cardoso:2005vk,Dias:2006zv}).  On the other hand, we were unable to find scalar field and Maxwell perturbations that are superradiant unstable (see appendix~\ref{app:ScalarMaxwell}), and the scalar field instability is similarly absent in the single spinning Myers-Perry black string \cite{Cardoso:2005vk}.  Furthermore, the lack of stable particle orbits about five dimensional equal-spinning Myers-Perry black holes suggests that the instability in the associated black string cannot exist in the eikonal ($m\to\infty$) limit. In spite of these less optimistic hints, we did find what we call {\it `finite-$m$' superradiant instabilities} in the tensor harmonic sector of gravitational perturbations. Consistent with the lack of particle orbits, this instability shuts-down in the eikonal limit, unlike known (uncharged) superradiant instabilities in black holes (see also discussion in footnote~\ref{foot:massiveScalar}). Here, the scalar field system is thus not a good toy model for the gravitational system (unlike in previously studied superradiant systems for black holes).

Let us now examine the gravitational superradiant instability in the equal-spinning Myers-Perry black string in more detail in light of our results and the above arguments.   We have found that this instability is generally bounded by extremality and two curves with $\mathrm{Im}\,\omega=0$.  One of these curves (the superradiant onset curve) satisfies $\omega=2m\Omega_H$, and follows the standard criterion for the onset of superradiance required for the amplification of waves (the factor of 2 reflects the fact that we have equal rotation along the two planes; see footnote~\ref{footCanPsi}).  However, the second curve with $\mathrm{Im}\,\omega=0$ (that we now call the confining cutoff curve) satisfies $\omega=k$, which suppresses the exponential falloff of fields, $e^{\pm \sqrt{k^2-\omega^2}\,r}$, and stops the superradiant instability due to a lack of confinement. That is, at this cutoff curve, the Schr\"odinger potential barrier that confines bound states ceases to exist.  Ultimately, in the eikonal limit, both of these mechanisms conspire to eliminate the superradiant instability.

As far as we are aware, this  mechanism of cutting off confinement is absent in other gravitational superradiant instabilities in black hole systems, such as in those with a potential barrier sourced by a massive field. The prototypical example is a massive scalar field in a Kerr black hole \cite{Zouros:1979iw,PhysRevD.22.2323,Dolan:2007mj}. In this case, superradiant instabilities exist as long as the mass of the scalar field perturbation is below an upper bound. But there is no (non-zero) lower bound for the mass of the scalar field for existence of superradiant bound states. In our case, the momentum of the perturbation along the extended black string direction provides an effective mass term to the system. Like the massive scalar on Kerr, we still find a maximum effective mass (i.e. momentum $k_{(0)}^{(m)} r_+$) for the existence of an unstable superradiant bound states. But, surprisingly, we also find a non-zero lower bound for this effective mass (i.e. the momentum $k_{\star}^{(m)} r_+$) below which we no longer have bound states and thus the system becomes stable to superradiance. 

How general is the finite-$m$ superradiant instability that we found? What circumstances allow this phenomenon to occur? Our superradiant gravitational instability of the equal-spinning Myers-Perry black string should also be present in the Kerr black string, though the Kerr case should instead be a more typical 'infinite-$m$' superradiant instability as it contains bound particle orbits.  Although a detailed study of perturbations in the Kerr black string requires solving PDEs, not ODEs. However, we argue around \eqref{Vinf} and in appendix~\ref{app:higherD} that it should not be present in $D>6$ equal-spinning Myers-Perry strings, so $D=5$ and $D=6$ strings seem to be very special cases.   What is also clear is that the gravitational superradiant instability of the six-dimensional equal-spinning Myers-Perry black string does not extend to scalar field nor Maxwell perturbations, even in $D=6$ (see appendix~\ref{app:ScalarMaxwell}). Moreover, in the single-spinning case, \cite{Cardoso:2005vk} argues that the effective potential of the scalar field also fails to allow for bound states.  Away from the eikonal limit, the status of superradiant instabilities in the single-spinning or more generally-spinning case remains unclear, and will again require PDEs to study in detail.

Our linear instability should have important consequences for the phase diagram of rotating black strings. Indeed, our linear results suggest that novel rotating black string solutions, that break the same rotational $U(1)_\psi$ isometry as our linear perturbation, should bifurcate from the onset of our superradiant instability. That is, new solutions with fewer isometries (those of our linear perturbation) than the Myers-Perry black string should exist. Fortunately, for $m=2$, our perturbations still preserve the $SU(2)$ isometry of the equal-spinning Myers-Perry solution, which indicates that these new solutions should be feasible to construct at the full nonlinear level. Actually, in section~\ref{sec:Superradiance3}, we argued that the onset of our $m=2$ superradiant instability should be a degenerate bifurcation line to a new two-parameter family of black string solutions. In general, the $U(1)_Z$ isometry of the system is broken, while this isometry is preserved for a special family. In two companion manuscripts, we will confirm this is the case. We will solve the full nonlinear problem (perturbatively to higher order and numerically) and find cohomogeneity-2 \textit{black resonator strings} with isometry group $\mathbb{R}_T\times SU(2)$ \cite{Dias:2022str}, and cohomogeneity-1 \textit{helical black strings} with isometry group $\mathbb{R}_T \times SU(2)\times U(1)_Z$ \cite{Dias:2023nbj}, where the coordinates $T,Z$ are defined in \eqref{rotcoord}.

Finally, let us comment on the endpoint of equal-spinning Myers-Perry black strings that are afflicted by one or both (Gregory-Laflamme and superradiant) instabilities.  There have only been a limited number of studies of the evolution of the Gregory-Laflamme instability in black strings, and all for the non-rotating case~\cite{Horowitz:2001cz,Lehner:2010pn,Emparan:2015gva,Figueras:2022zkg} (see also \cite{Figueras:2015hkb} for the evolution of the Gregory-Laflamme instability of black rings). Thus far, these results are consistent with entropic expectations.  If non-uniform string solutions are entropically preferred, which is the case for high spacetime dimensions~\cite{Gubser:2001ac,Wiseman:2002zc,Sorkin:2004qq,Kudoh:2004hs,Sorkin:2006wp,Headrick:2009pv,Figueras:2012xj}, then they provide a natural endpoint~\cite{Emparan:2015gva}.  Otherwise, evolution proceeds to a ``localised'' black hole configuration with spherical horizon topology, and the evolution process goes through a horizon pinch-off phase that necessarily violates the weak cosmic censorship conjecture~\cite{Lehner:2010pn,Figueras:2022zkg}.  Of course, classical evolution stops to be valid before reaching the naked singularity, so the ultimate classical endpoint as predicted by entropy arguments requires that the naked singularity be resolved in a sufficiently mild way. The evolution of the Gregory-Laflamme instability for the rotating black string should proceed in a similar way. That is, that the most entropic solution is the preferred endpoint, whether this may be some non-uniform string or localized black hole.\footnote{Note that it is not yet known whether the $D=6$ non-uniform equal-spinning black strings of \cite{Kleihaus:2007dg} are entropically favored over the uniform strings.
}

What about the endpoint of equal-spinning Myers-Perry black strings that are only unstable to superradiance?
The time evolution of the gravitational superradiant instability is best understood for small rotating black holes in global anti de-Sitter (AdS) space.  In this context, it was argued that the unstable Kerr-AdS black holes develop towards `{\it black resonators}', which are black holes (of pure Einstein-AdS gravity) with gravitational hair that have a single Killing vector field (the horizon generator). That is, they are not time independent nor axisymmetric, but are time-periodic \cite{Dias:2015rxy} (see \cite{Dias:2011at,Dias:2011tj,Dias:2015rxy,Ishii:2018oms,Ishii:2020muv,Ishii:2021xmn} for more details).  However, black resonators are themselves unstable to superradiant instabilities with arbitrarily higher azimuthal numbers $m$ (unlike the case for Myers-Perry black strings).
The ultimate endpoint remains an open problem but if, as conjectured in \cite{Dias:2011at,Cardoso:2013pza,Dias:2015rxy,Niehoff:2015oga}, the cascade continues to ever increasing $m$'s, then a significant amount of energy will be pushed towards microscopic length scales, violating the weak cosmic censorship conjecture \cite{Dias:2015rxy,Niehoff:2015oga}.  The numerical time evolution of this instability is currently consistent with this picture \cite{Chesler:2018txn,Chesler:2021ehz}.

Note that arguments for the existence of this cascade largely comes from the fact that there are an infinite number of unstable modes with arbitrarily high azimuthal numbers $m$ affecting the original black hole superradiant system.  The superradiant instability for the equal-spinning Myers-Perry black string does not have this property. Indeed, a given Myers-Perry black string can be unstable to only a few modes with different $m$'s (not an infinite tower of them, unlike Kerr-AdS). This is ultimately related to the fact that the finite-$m$ superradiant instability shuts down in the eikonal limit.  For this reason, it is expected that no continuous cascade to higher $m$'s should be present in the time evolution of the superradiant instability of the Myers-Perry black string.  Rather, it seems more plausible that the superradiant instability will just drive the solution to a particular black resonator string or helical string with a given $m$, whichever has the most entropy at the same energy $\mathcal{E}$ and angular momenta $\mathcal{J}$.

Thus far, we have discussed the Gregory-Laflamme and superradiant instabilities in isolation, but as we have shown, both instabilities can be present and thus compete for certain parameters of the Myers-Perry black string.  The typical growth rate for the Gregory-Laflamme instability (in units of the horizon radius) is $10^{-1}$ while that of the superrradiant instability is $10^{-5}$.  This large separation is not surprising, given the vastly different origins of the instabilities.  One is governed primarily by horizon dynamics while the other is driven by the amplification and confinement of gravitons, and these typically have very different timescales.  Given the disparity of growth rates, it should be the case that the Gregory-Laflamme instability will likely dominate the early dynamics in cases where both instabilities are present.  We stress, though, that our linear analysis will quickly cease to be valid in mid to late times in the evolution. It is also unclear if one instability remains present even when the other has settled down.  We do not know, for instance, if rotating localised Kaluza-Klein black holes have a superradiant instability (although the typical arguments for superradiance suggest that they should). There will be various interplay of the Gregory-Laflamme and superradiant instabilities, which would be interesting to address in future work.

\begin{acknowledgments}

The authors would like to thank Roberto Emparan, Takahisa Igata, Akihiro Ishibashi, and Masashi Kimura for useful discussions.
O.J.C.D. acknowledges financial support from the  STFC ``Particle Physics Grants Panel (PPGP) 2018" Grant No.~ST/T000775/1. OD acknowledges the Isaac Newton Institute, Cambridge, and the organizers of its long term programme ``Applicable resurgent asymptotics: towards a universal theory" during which this work was completed.
The work of T.I. was supported in part by JSPS KAKENHI Grant Number 19K03871.
The work of K.M. was supported in part by JSPS KAKENHI Grant Nos. 20K03976, 21H05186 and 22H01217.
B.W. acknowledges support from ERC Advanced Grant GravBHs-692951 and MEC grant FPA2016-76005-C2-2-P. J.~E.~S. has been partially supported by STFC consolidated grant ST/T000694/1. The authors also acknowledge the use of the IRIDIS High Performance Computing Facility, and associated support services at the University of Southampton, in the completion of this work.
\end{acknowledgments}

\appendix

\section{Scalar field and Maxwell perturbations}\label{app:ScalarMaxwell}

As described in the main text, we also attempted to find, numerically, superradiant instabilities in the sector of scalar field and Maxwell perturbations about the equal-spinning Myers-Perry black string. But, unlike for the gravitational sector, we have found no superradiant instability in these sectors of perturbations. To complement this negative search, in this appendix, we give analytical arguments in favour of the absence of superradiant instability for scalar field and Maxwell perturbations. These arguments are similar to those employed in section~\ref{sec:Superradiance2} $-$ in the discussion between \eqref{gammaGrav}-\eqref{OmegaChighm} $-$ to argue in favour of the gravitational instability. Here we show only the calculations for massless fields; for massive fields the analysis and conclusion are similar.\footnote{The mass $\mu$ can be introduced simply by replacing $k^2 \to k^2 + \mu^2$ in our analysis, while the final results for the condition of ${\Omega}_H^{(m)}/\Omega^{\mathrm{ext}}_H$ are not altered. For gravitational perturbations in the main text, a graviton mass would affect the value of $k$ for the instability, but we are interested only in Einstein gravity.}

\subsection{Scalar field}
Consider the Klein-Gordon equation, $\Box \Pi=0$, for a massless scalar field $\Pi$ on the equal-spinning Myers-Perry black string background. Introducing the separation ansatz
\begin{equation}\label{AppScalarPerturbation}
\Pi(t,r,\theta,\phi,\psi,z) = e^{\mathrm{i}kz}e^{-\mathrm{i}\omega t} \Pi (r) D^j_{\ell'  \ell=-m}(\theta,\phi,\psi)\,,
\end{equation}
where $D^j_{\ell'  \ell=-m}$ are the Wigner D-matrices introduced in section~\ref{sec:Wigner}, the Klein-Gordon equation  decouples for each $m$ and the radial equation one needs to solve reads
\begin{equation}\label{AppKGeq}
\left( r^3 F \Pi' \right)' + r \left[ \frac{H r^2}{F}\left(\omega -2 m \frac{\Omega}{H}\right)^2 -\frac{4 m^2}{H}-4(j^2+j-m^2)- k^2 r^2 \right] \Pi = 0\,.
\end{equation}
Defining $\Phi = (r H)^{1/4}\Pi$, we can rewrite this equation in the Schr\"{o}dinger form \eqref{Scheq} where in the present case the effective potential is
\begin{equation}
 V = \frac{F}{H} k^2 - \left(\omega -2 m \frac{\Omega}{H}\right)^2 + \frac{4 F}{r^2 H} \left[ j(j+1) - m^2 \left(1 - \frac{1}{H}\right)\right]
+\frac{1}{r^{3/2} H^{1/4}} \frac{\mathrm{d}^2}{\mathrm{d} r_\ast^2}(r^{3/2} H^{1/4})\,.
\end{equation}
Near infinity, this Schr\"{o}dinger  potential reads
\begin{equation}
 V|_{r\to \infty}\sim k^2-\omega^2+\left[4j(j+1)+\frac{3}{4}-\frac{4k^2r_+^2}{4-\Omega_H^2r_+^2}\right]\frac{1}{r^2}+ \mathcal{O}(r^{-4})\, .
\end{equation}
Alike for the gravitational perturbation~\eqref{Vinf}, the scalar field perturbation can be confined near the horizon for a sufficiently large $k$.
Yet, unlike the gravitational case, a simple analytical analysis of our radial equation suggests absence of scalar field superradiant instabilities as described next.

At infinity, the asymptotic behaviour of the field is given in general by $\Pi|_{r\to \infty} \sim r^{-3/2} e^{\pm\sqrt{k^2-\omega^2}\,r}$. But for $\omega = k$, the exponential decay is absent and we have to follow an analysis similar to the one described in \eqref{gammaGrav}-\eqref{OmegaChighm} for the gravitational sector. Namely, for $\omega = k$, the asymptotic behaviour of the scalar field is instead $\Pi|_{r\to \infty} \sim r^{-1 \pm \gamma}$ with $\gamma = \sqrt{(2j+1)^2 -4\widetilde{k}^2/(4-\widetilde{\Omega}_H^2)}$. As for the gravitational case, the special $\gamma=0$ degenerate case should select the confining cutoff condition $\widetilde{k}^{(m)}_{\star}(\widetilde{\Omega}_H)$ for the scalar superradiant instability. So setting $\gamma=0$, we obtain
\begin{equation}\label{AppCutoffScalar}
\widetilde{k}^{(m)}_{\star}=\frac{1}{2}(2j+1)\sqrt{4-\widetilde{\Omega}_H^2}\,.
\end{equation}
So this should define the scalar field counterpart of the superradiant cutoff line $cB$ of the gravitational case observed in Fig.~\ref{Fig:zeroMode}. Actually, we can also proceed and apply the same argument leading to \eqref{OmegaChighm} to get the scalar field counterpart of the critical point $c$ where the onset curve for $\widetilde{k}^{(m)}_{(0)}(\widetilde{\Omega}_H)$ and cutoff curve  $\widetilde{k}^{(m)}_{\star}(\widetilde{\Omega}_H)$ should intersect. Recall from Fig.~\ref{Fig:zeroMode}, that $c$ is a vertex point for the superradiant unstable region, i.e. instability is present only for rotation above the one of this point.
In this context, the onset of the scalar superradiant instability must certainly satisfy the condition $\omega = 2 m \Omega_H$ (this follows from the equation of motion at the horizon). Combining this with $\omega = k^{(m)}_{\star}$ given in \eqref{AppCutoffScalar}, we find that the scalar field superradiant instability should only exist for horizon angular velocities higher than the critical value
\begin{equation}\label{AppOmegaCritScalar}
\frac{{\Omega}_H^{(m)}}{\Omega^{\mathrm{ext}}_H}=\frac{\sqrt{2}(2j+1)}{\sqrt{(2j+1)^2 + 4m^2}} > 1\,,
\end{equation}
where the last inequality follows from $|m| \le j$, and the ratio approaches $1$ from above as $m \to \infty$. So the minimum angular velocity required for the scalar field instability is larger than the velocity at extremality. Effectively, this argument rules out the existence of a superradiant instability in the scalar field sector. Our arguments did rely on some assumptions that are only known to be valid empirically for the gravitational sector.  But, as stated above, we have also searched directly for instabilities of the scalar field by solving the scalar field eigenvalue problem numerically but found no instability.

\subsection{Maxwell field}
For the Maxwell field, we focus on the perturbation given by
\begin{equation}\label{AppMaxwellPerturbation}
a_M \mathrm{d}x^M = e^{\mathrm{i}kz}e^{-\mathrm{i}\omega t} A (r) D^j_{\ell' \ell=-j}(\theta,\phi,\psi) \sigma_- \,.
\end{equation}
where $D^j_{\ell'  \ell=-m}$ are again the Wigner D-matrices introduced in section ~\ref{sec:Wigner}.
The azimuthal mode number for this perturbation is $m=j+1$ and it is a half-integer multiple satisfying $m\geq 1$.
This mode decouples from other Maxwell perturbations and it is expected to be the most superradiant unstable mode for a given $j$.
The gauge transformation of the Maxwell field is $a_M\to a_M + \partial_M \lambda$.
The gauge parameter $\lambda$ is also expanded by the Wigner D-matrices as $\lambda=e^{\mathrm{i}kz}e^{-\mathrm{i}\omega t} \Lambda (r) D^j_{\ell' \ell}$. The maximal azimuthal mode number of the gauge parameter $\lambda$ is $m=j$ for a given $j$. It follows that the Maxwell perturbation~(\ref{AppMaxwellPerturbation}) which has $m=j+1$ is invariant under the gauge transformation.

The Maxwell equation $\Box a_M = 0$ takes the form
\begin{equation}\label{AppMaxwelleq}
\left( rFA' \right)' + \frac{1}{r} \left[ \frac{H r^2}{F}\left(\omega -2 m \frac{\Omega}{H}\right)^2 -\frac{4 m^2}{H}- k^2 r^2 \right] A = 0\,.
\end{equation}
Defining $\Phi = r^{1/2} H^{1/4}A$, we recast this as a Schr\"{o}dinger equation \eqref{Scheq} with the effective potential
\begin{equation}
 V = \frac{F}{H} k^2 - \left(\omega -2 m \frac{\Omega}{H}\right)^2 + \frac{F}{r^2 H} \left[\frac{4m^2}{H}-(rF)'\right]
+\frac{1}{r^{3/2} H^{1/4}} \frac{\mathrm{d}^2}{\mathrm{d} r_\ast^2}(r^{3/2} H^{1/4})\,.
\end{equation}
Its asymptotic behaviour is
\begin{equation}
 V|_{r=\infty}  \sim k^2-\omega^2+\left(4m^2-\frac{1}{4}-\frac{4k^2r_+^2}{4-\Omega_H^2r_+^2}\right)\frac{1}{r^2}+ \mathcal{O}(r^{-4})\,,
\end{equation}
which indicates that Maxwell perturbations can be confined for a sufficiently large $k$.  However, there is no room for the existence of a superradiant instability as argued next (following again a line of reasoning similar to the one employed in the gravitational discussion of \eqref{gammaGrav}-\eqref{OmegaChighm}).

Typically, Maxwell perturbations decay exponentially as $A|_{r=\infty} \sim r^{-1/2} e^{\pm\sqrt{k^2-\omega^2}\,r}$. But for $\omega = k$, the exponential decay is absent and the asymptotic behaviour of the Maxwell field is instead $A|_{r\to \infty} \sim r^{\pm \gamma}$ with $\gamma = \sqrt{4m^2 - 4\widetilde{k}^2/(4-\widetilde{\Omega}_H^2)}$. As for the gravitational case, the special $\gamma=0$ degenerate case should select the confining cutoff condition $\widetilde{k}^{(m)}_{\star}(\widetilde{\Omega}_H)$ for the Maxwell superradiant instability. This $\gamma=0$ condition yields
\begin{equation}\label{AppCutoffMax}
\widetilde{k}^{(m)}_{\star} =m\sqrt{4-\widetilde{\Omega}_H^2}\,,
\end{equation}
which should define the Maxwell field counterpart of the superradiant cutoff line $cB$ of the gravitational case observed in Fig.~\ref{Fig:zeroMode}. Very much like for the gravitational discussion leading into \eqref{OmegaChighm}, the endpoint $c$ of this cutoff curve can be obtained from the condition that the cutoff curve \eqref{AppCutoffMax} intersects the onset curve $\widetilde{k}^{(m)}_{(0)}(\widetilde{\Omega}_H)$ with  $\omega = m \Omega$. This condition concludes that the superradiant instability for Maxwell fields should only exist for horizon angular velocities higher than the critical value
\begin{equation}\label{AppOmegaCritMax}
\frac{{\Omega}_H^{(m)}}{\Omega^{\mathrm{ext}}_H} = 1\,,
\end{equation}
i.e.~for equal-spinning Myers-Perry strings rotating with a velocity above the extremal value (which are not regular).
Thus non-extreme Myers-Perry black strings do not develop superradiant instabilities for the Maxwell field. This conclusion is supported by the fact that we directly searched numerically for instabilities but found none.

We can repeat the same analysis for massive scalar and Proca fields and obtain the same conclusion that these fields do not induce superradiant instability. This suggests that the mass of the fields is not essential, and spin dependence is important for the existence of the superradiant instability for Myers-Perry black strings.

\section{Absence of superradiant instability in equal-spinning Myers-Perry strings in \texorpdfstring{$D>6$}{}\label{app:higherD}}

In the main text we found that $D=6$ equal angular momenta Myers-Perry black strings are unstable to the finite-$m$ superradiant instability (in addition to being unstable to the Gregory-Laflamme instability). In particular, we also found a criterion $-$ detailed in the discussion of \eqref{gammaGrav}-\eqref{OmegaChighm} of section~\ref{sec:Superradiance2} $-$ to find, almost only resorting to analytical computations, the superradiant unstable region $ABc$ in the phase diagrams of Fig.~\ref{Fig:zeroMode} and Fig.~\ref{fig:stabilityDiag}.
We also argued that $D=5$ Kerr black strings should also be superradiant unstable since it has superradiant bound states. However, in this appendix we employ a line of reasoning that follows the discussion of \eqref{gammaGrav}-\eqref{OmegaChighm} of section~\ref{sec:Superradiance2} to argue that superradiant instabilities should not be present for $D>6$ equal angular momenta Myers-Perry black strings. In short, although there are superradiant modes, there is no potential that can confine them as bound states in a region of the parameter space that is contained inside the boundary set by the extremal (zero temperature) Myers-Perry black string configuration. So the $D=6$ case studied in the main text is very special.

Consider $D=2N+4$ dimensional equal-spinning Myers-Perry black strings, i.e. the product of an equal-spinning $(2N+3)$-dimensional Myers-Perry black hole and a circle.
Unlike for the $N=1$ case, we argue that the gravitational superradiant instability is absent for $N\geq 2$.
The metric of the equal-spinning Myers-Perry black string in $D=2N+4$ dimensions is
\begin{equation}
{\mathrm d}s^2 _{\rm MP \,string}= -\frac{F}{H}\,{\mathrm d}t^2 +\frac{{\mathrm d}r^2}{F} + r^2 \left[ H\left(d\psi+A_a dx^a -\frac{\Omega}{H}\, {\mathrm d}t \right)^2+ \hat{g}_{ab}dx^a dx^b  \right] + \mathrm dz^2\,,
\end{equation}
where
\begin{equation}
F(r)= 1-\frac{r_0^{2N}}{r^{2N}}+\frac{a^2r_0^{2N}}{r^{2N+2}}\,, \qquad H(r)=1+ \frac{a^2 r_0^{2N}}{r^{2N+2}}\,, \qquad
\Omega(r)=\frac{a \,r_0^{2N}}{r^{2N+2}}\, ,
\end{equation}
and $\hat{g}_{ab}$ is the Fubini-Study metric on $CP^N$ and $A_a$ is the K\"{a}hler potential.
In \cite{Kunduri:2006qa}, ``doubly transverse'' gravitational perturbations of the equal-spinning Myers-Perry black hole have been considered for $D=2N+3$ dimensions with $N \ge 2$. A unified form of the perturbation equation was given for decoupled gravitational and scalar field perturbations. While their study was limited to $N \ge 2$, the corresponding perturbation was examined in detail for $N=1$ in \cite{Murata:2008yx}. Below we will check that the unified form of \cite{Kunduri:2006qa} also applies to $N=1$.

The scalar-gravitational unified treatment of \cite{Kunduri:2006qa} can be generalized straightforwardly to the $D=2N+4$ dimensional Myers-Perry black string with $N \ge 1$ since one just needs to add an extra dimension to the formalism which introduces a Kaluza-Klein mass term to the perturbation equations. The unified perturbation equation can be written in the Schr\"odinger form
\begin{equation}\label{UnifiedPertEq_rcoord}
-\frac{\mathrm{d}^2}{\mathrm{d} r_\ast^2} \Phi + V \Psi = 0\,,
\end{equation}
where $dr_\ast=\sqrt{H}/F dr$ denotes the tortoise coordinate.
The unified Schr\"odinger potential for scalar field and anti-hermitian gravitational perturbations is given by
\begin{align}
V = &\frac{F}{H} k^2 - \left(\omega-2 m \frac{\Omega}{H}\right)^2 + \frac{4 F}{r^2 H} \left[ l(l+N) - m^2 \left(1 - \frac{1}{H}\right)\right] \nonumber\\
& +\frac{1}{r^{N+1/2} H^{1/4}} \frac{\mathrm{d}^2}{\mathrm{d} r_\ast^2}\left(r^{N+1/2} H^{1/4}\right)\,,
\label{VhighD}
\end{align}
where $l=n+|m| \ (n=0,1,2,\dots)$ for the scalar field, while $l$ is restricted to $l \ge 1$ for the gravitational perturbation, and $m$ takes either integer or half-integer values.\footnote{The quantum numbers in \cite{Kunduri:2006qa} are related to ours as $m_\mathrm{theirs} = 2 m_\mathrm{ours}$ and $l_\mathrm{theirs} = 2 l_\mathrm{ours}$. Also, the perturbations we focus on corresponds to ``$\sigma=1$'' in \cite{Kunduri:2006qa}.}
We have checked that \eqref{UnifiedPertEq_rcoord} reproduces the perturbation equation we have been using for $N=1$. Indeed, setting $l=j$, the scalar field perturbation \eqref{AppKGeq} is reproduced. Moreover, for $l=m-1$ one recovers \eqref{EOMsuper}.\footnote{Also, the range of $l$ is consistent. In \cite{Kunduri:2006qa}, the range of $l_\mathrm{theirs}$ was restricted to $l_\mathrm{theirs} \ge 2$, which would also applied to the case of $N=1$. This condition is translated to $l \ge 1$ for our $l$. Meanwhile, we have $m \ge 2$, which gives $l=m-1\ge 1$.}

For $N\geq 2$, the asymptotic behaviour of the potential $V$ becomes
\begin{equation}
 V|_{r\to \infty} \simeq k^2-\omega^2 +\frac{4l(l+N)+N^2-1/4}{r^2} + \mathcal{O}(r^{-4})\, .
\label{VinfhighD}
\end{equation}
Unlike for the $N=1$ case of \eqref{Vinf}, the $1/r^2$ term is always positive.
Moreover, because  $F/H|_{r\to \infty} \simeq 1-r_0^{2N}/r^{2N}$  in \eqref{VhighD}, $k^2$ does not appear in the coefficient of the $1/r^2$.
Altogether, this suggests that for $N\geq 2$, the scalar field and gravitational perturbations cannot be confined inside the ergoregion and thus we should not expect the existence of a superradiant instability.
We actually searched numerically for the onset of the instability for $N=2,3,4$, but we did not find evidence of its existence, consistent with the above analysis.

We can also repeat the same line of arguments  detailed in the discussion of \eqref{gammaGrav}-\eqref{OmegaChighm} of section~\ref{sec:Superradiance2} (and further explored in appendix~\ref{app:ScalarMaxwell}) to arrive to the same conclusion.
For $k\neq\omega$, the asymptotic solutions of \eqref{UnifiedPertEq_rcoord} decay exponentially as $\Psi|_{r\to \infty} \sim e^{-\sqrt{k^2-\omega^2}\, r}$. However, for $k=\omega$, the two possible decays are instead polynomial. Concretely, for $N \ge 2$, they read
\begin{equation}\label{gammaGravHighN}
\Psi \sim r^{\frac{1}{2} \pm (l+N)}\,.
\end{equation}
Unlike the $N=1$ counterpart case discussed around \eqref{gammaGrav},
there is no room for the two solutions of \eqref{gammaGravHighN} to become degenerate since $l+N>0$ (and more importantly the exponents are not a function of $k$ and $\Omega_H$). Hence, this analysis also gives no evidence for the existence of a superradiant instability (be it in the scalar or gravitational sector) for $N \ge 2$ ($D\geq 8$) equal-spinning Myers-Perry black strings, unlike for the $N=1$ ($D=6$) case discussed in the main text.

\bibliography{refsResonator}{}
\bibliographystyle{JHEP}

\end{document}